\documentclass{article}
\usepackage[utf8]{inputenc}
\usepackage[a4paper, total={6in, 8in}]{geometry}
\usepackage{fullpage}
\usepackage{float}
\usepackage{subfigure}
\usepackage{soul}
\usepackage{amsmath}
\usepackage{enumitem}
\usepackage{pdfpages}
\usepackage{dingbat}
\usepackage{pifont}
\usepackage{graphicx}
\newcommand{\xmark}{\ding{55}}
\usepackage{tabularx}
\usepackage{color, colortbl}
\usepackage{url}
\usepackage{comment}

\usepackage{hyperref}
\hypersetup{
    colorlinks=true, %set true if you want colored links
    linktoc=all,     %set to all if you want both sections and subsections linked
    linkcolor=blue,  %choose some color if you want links to stand out
}
\usepackage{authblk}
\providecommand{\keywords}[1]
{
  \textbf{\textit{Keywords:}} #1
}

\title{A Technical Look At The Indian Personal Data Protection Bill}
\author[1]{Ram Govind Singh
\thanks{Indian Computer Emergency Response Team (CERT-In), India}}
\author[1,2]{Sushmita Ruj}

\affil[1]{Indian Statistical Institute Kolkata, India \protect\\
\texttt{Ramgovind.2010@gmail.com, sush@isical.ac.in}}
\affil[2]{CSIRO Data61, Australia \protect\\ \texttt{Sushmita.Ruj@data61.csiro.au}}
\date{}

\begin{document}
% \author{Ram Govind Singh\inst{1} \and Sushmita Ruj\inst{2}}
% %\authorrunning{R.~Singh and S.~Ruj}
% \institute{Indian Computer Emergency Response Team (CERT-In),\\
% 	\email{ramgovind.2010@gmail.com}
% 	\and
% 	Data61, CSIRO, Australia,\\
% 	\email{Sushmita.Ruj@data61.csiro.au}}

\maketitle

\begin{abstract}
  The Indian Personal Data  Protection  Bill 2019 provides a legal  framework  for  protecting  personal  data. It is modeled after the European Union's  General Data Protection Regulation(GDPR). We present a detailed description of the Bill, the differences with GDPR, the challenges and limitations in implementing it. We look at the technical aspects of the bill and suggest ways to address the different clauses of the bill. We mostly explore cryptographic solutions for implementing the bill. There are two broad outcomes of this study. Firstly, we show that better technical understanding of privacy is important to clearly define the clauses of the bill. Secondly,  we also show how technical and legal solutions can be used together to enforce the bill.  
\end{abstract}
%\hspace{2cm}   

\newpage
\tableofcontents
\newpage

\section{Introduction}
 \textit{Privacy} (also called as data privacy or information privacy) is defined as the ability of an individual or an organization to decide when, whom, and how much data in a computer system may be disclosed to a third party. It is a \textit{fundamental right}  in the Indian constitution \cite{privacy}. The \textit{fundamental rights} are a set of rights which requires a higher degree of protection from the government irrespective of a person's caste, race, religion, gender or place of birth.  ``Any individual, group, or organization can exclude themselves or information about themselves, thereby, representing themselves selectively".

%Privacy is the right and ability of choice and preferences. Privacy is a  fundamental right  in the Indian constitution \cite{privacy}.  Fundamental rights are a set of rights which requires a higher degree of protection from the government irrespective of a person's caste, race, religion, gender or place of birth. Any individual, group, organization or companies may exclude themselves or information about themselves thereby represent themselves selectively. Hence, in a digital era too, the privacy (also called as data privacy or information privacy) is defined as the ability of an individual or an organization to decide when, whom and how much data in a computer system may be disclosed to a third party. Data privacy  is expected to be preserved by any individual or organization while dissemination of data and technology. 
\par
Data privacy  is expected to be preserved by any individual or organization while disseminating the data and technology.  Typically, an organisation collect data from a small amount to a large extent with/without user’s consent. This data may be used for purposeful processing as well as for other purposes like data analysis,
marketing, automated decision making and profiling. An organisation may share this data with third parties also without prior knowledge of data owners. Further, the data which are not protected properly
by a system may lead to various risks like location disclosure, profile disclosure, data traffic analysis, behaviour monitoring or identity disclosure. 

The goal of data protection is to identify the possibilities
of these risks and to develop proper technologies and services in such a way which can eliminate such risks \cite{jio}. In particular, data protection concern about the protection of data from breaches, disclosure, modification and privacy leakage throughout the life cycle of data. Although the organizations are using various technologies to  protect data like encryption, authentication, authorization, access control etc, however, there are still so many data breaches happening around the world \cite{jio,Facebook, AndriodApp}. Even if the data breach is protected, it does not ensure the privacy of data. For instance, consider an online shopping website where a user, after login, did shopping and paid online. If  the merchant is not trusted, the personal data may be disclosed to third parties without the user's awareness. Such data can be sold, distributed or shared with  other parties for monetary purpose  \cite{Facebook,TrueCaller,LASues}. Therefore, \textit{the assurance of data protection in the current system  does not ensure data is indeed protected, and  privacy is preserved}. So, we require two components: first, the law and obligation under which processing will be carried out; second,  a set of technical method that can translate legal tenets into the technical framework.
%Presently data owner does not know  how, where, and in which form, their data has stored and processed. Who will take care of users right if organisations are using the data maliciously. 
Here, the law and regulation will ensure legal binding and penalties for data processing while the technical standard  will ensure security, safety and privacy of data.  
\begin{equation*}
\mbox{Data Protection} = \mbox{Legal Framework} + \mbox{Technical Standards}
\end{equation*}

%\textbf{Data protection= data protection(using technical standard)  +  Consent(Data owner preferences) + Data protection framework(Regulation and Law for transparency integrity and security)} 

Many countries have introduced a data protection framework, or some other  kind of legislation  to protect user's data \cite{dp}. European Union(EU) is the first one who has proposed General Data Protection Regulation(GDPR) \cite{GDPR} that will work as a legal tenet to protect users data within EU territory. As per GDPR, it is the data processor's responsibility to protect the user's data. GDPR provides several rights to data owner over their data, impose regulation on processing,  and restriction over organizations how they can use personal data. Similarly, Government of India has recently proposed the Personal Data Protection Bill-2019 (PDPB-2019) \cite{IDPL} to provide legal framework for protecting personal data from organizations who are operating their business within India. It is a draft  to protect the data and privacy of all individuals residing within the territory of India, and describes how data processing shall be outside India. As per the framework, now companies have to modify their data processing activity, privacy policy, and cookies policy in such a way that is compliant with the framework.
In order to be compliant various technical and engineering tools need to be in place. It is important to take preventive measures to protect the privacy of individuals, as much as it is important to take legal actions after violations of privacy. In other words technical implementation and challenges should be considered while formulating the requirements of such a policy and formulate the legal guidelines.  

Our aim is two fold: First, to understand the problem and definitions of privacy and analyse the bill in light of these definitions. Second, to provide technical guidelines to organizations to make data processing and storage compliant with the legal framework. 
%In doing so,  suggest how the  bill can be implemented using technical and engineering solutions. An analogy is  Second, to study the technical and engineering challenges and provide guidelines to enforce the   This not only helps to formulate legal guidelines, but also provides a roadmaps to organizations to with the Legal enforcement On the contrary, to achieve various features of it,  we need to implement  it technologically. This will require  more advanced cryptographic schemes, technology and algorithmic solutions  to  fulfil the criteria of the framework. \\

\subsection{Motivation} To the best of our knowledge, no work is done in the context of technical implementation of PDPB.  In the area of GDPR, the work has been  done mostly in the five major categories: (i) on the challenges and limitations of GDPR \cite{BigDataPbD,Onpurpose,BuisnessOfPersonalData,DigitalPrivacyAFailure}; (ii) on the properties of GDPR  like right to forget,  and how it can be achieved \cite{BlockchainMutability,EfficacyOfGDPR}; (iii) how changes should be done in the current system that can fulfil the requirement of obligations of GDPR \cite{TakeSomeCookies,DPbD,BigDataPbD}; (iv) the possible architecture of the system as per GDPR standard and (v)design and implementation of consent. Since there are few differences between GDPR and PDPB, some of the above work cannot be trivially applied for PDPB. Also, there are still various categories where technological aspects needs to be explored in GDPR. We therefore ask the following questions:

\begin{quote}
    \begin{enumerate}
        \item What issues do the current bill address? 
        \item What are the properties of obligations of various entities that will need technical implementation?
        \item What are the various technologies that might be required to satisfy the conditions given in the bill?
        \item How can organizations implement the data management framework given in the bill?
        \item How to make it easier for individuals to verify that the terms and conditions are honoured.  
        \item How to make it easier for law enforcement agencies to validate claims, in the face of a dispute regarding the enforcement of the policies.
        \item In trying to address the above issues, we analyse the existing bill and suggestion modifications that we feel are  important to protect  data privacy. 
      % \item What kind of proof will be required by the data fiduciary and data principal in order to achieve consensus while implementing the properties of obligations?
        %\item How the above proof can be implemented?
    \end{enumerate}
     
\end{quote}

%But no work is done regarding description of technologies that may help to implement the rules of IDPL and GDPR. Also, no work is done on the analysis of properties that will be needed to implement it technologically. It is also need to define what kind of proof by data fiduciary and data principal will require to define and prove technologically in order to follow obligations. In addition to this, there is no work done on how this proof can be implemented. 

\noindent In this paper, we have solved the above problem.  We believe that  a better understanding of the technical challenges, makes it easier to formulate the bill and  guidelines to be followed to protect data privacy.  

%We have provided a detailed list of properties of obligations and argued that implementation of all properties will required in order to implement obligation. For each description of properties of obligation  we have given detailed  list of technological methods that can be used to  implement it. A description is provided for how legal obligations of this regulation will be translated into technical standards and can be implemented technologically in order to achieve all the goal of IDPL. For each obligation and its properties, we have provided set of either cryptology based solution or other kind of technical solution. Conversion of  tenets into technical framework will require to know the answer of  questions like,  how one will come to know that  an organization is not processing the data for purposes other than specified? How can be demonstrated  that data has collected after consent has obtained from data principal. Further, how can be ensured  that user's right like right to forget, right to access and right to correction has been implemented properly. Throughout the paper we have discussed that, data must be protected throughout the life cycle of data. To achieve this data fiduciary will require to give security proof of such obligations. \\

\subsection{Our Contribution} Our major contributions are as follows: 
  \begin{itemize}
      \item  A detailed description of Indian personal data protection bill (PDPB), its major components, and outline of various obligations is provided. 
      \item We have shown the differences of PDPB and GDPR with respect to regulation point of view. 
       
      \item For each major obligation of PDPB framework, we have mentioned all the properties, challenges, and security concern that need to be implemented in order to fulfil the requirement of obligations. 
      %Translation of legal obligation into technical implementation require assurance of   implementation of all these properties technologically. 
      A detailed analysis is given for various obligations such as data principal consent, data collection, data processing, security by design, transparency and data audit.
     
      \item For the above properties and challenges, we have described how cryptographic (such as encryption, signature schemes, zero-knowledge proof etc) and other  solutions (such as anonymisation, de-identification, access control, etc)  can be used for technical implementation. 
      %At various places, we have proved that cryptology  is not sufficient to full-fill all the goals that's why we shall require to incorporate some additional mechanism like  advanced data processing methods, identification and anonymization methods, access control, audit and blockchain based methods in order to build more strong model for data protection framework. 
       
      \item We have  mentioned the  set of challenges in obligations where existing solutions can not solve the problem. For such challenges, we  advanced methods have to be explored. 
       
      \item Using the example of the permanent account number (PAN), we show that  achieving the goal of data protection is a collective responsibility of multiple organization. If a single organization is modifying their business model as per the data protection framework, it will require other communicating parties too, to change their business model. 
       
     %  \item Further, we have proposed that blockchain can be used to build more strong solution of tenets. We have mentioned how existing development in blockchain can  be used to resolve dispute, to provide transparency and to build more stronger model of data protection  framework through its distributed nature.
       
      \item Lastly, we have provided the existing limitations of Indian data protection framework that needs to be modified.
  \end{itemize}
   
\subsection{Organization} The rest of the paper is organized as follows. Section 2 describes related work and section 3 contains background and description of PDPB. Section 4 discuss the comparison of PDPB and GDPR. Next, section 5 is about the major obligations of PDPB and their implementation. Further, we have provided the limitation of PDPB in section 6. At last, conclusion of the paper is provided. 
\section{Related work}

Privacy of user's personal data is very old concept. It was first proposed by Cavoukian in \cite{PrivacyByDesign}. Earlier development of privacy oriented application was, user's centric. The goal was application should be privacy oriented so that user's can control their privacy and personal data. 
The earliest privacy laws were formulated by France (1978) \cite{France_PA} and Canada(1983) \cite{canada_PA}.  Similarly Australia (1988)  \cite{Australia_PA}, New Zealand (1993) \cite{Newzealand_PA} and USA (HIPAA,1996) \cite{HIPAA}  have also formulated privacy acts that set guidelines to collect, use, disclose and share the personal information.
Many countries and economies have also come together to establish regulatory frameworks for cross border transfer of personal information. Such frameworks include  APEC Cross-Border Privacy Rules (CBPR) System \cite{APEC}, OECD Guidelines on the Protection of Privacy and Transborder Flows of Personal Data \cite{OECD} and EU-US shield framework \cite{USEU_shield_framework}.
%With the enforcement  of data protection regulations like GDPR \cite{GDPR} and HIPPA \cite{HIPAA}, 
%The protection of privacy and user's personal data is now the responsibility of third party entity : the \emph{data fiduciary}. 
PDPB \cite{IDPL} is also one step towards data protection. Currently very little has been discussed about PDPB.  Since PDPB has been inspired by GDPR and has a good deal of similarity,  we have described some of the earlier work done for GDPR.
\par
The biggest challenge in the design of the data protection framework is to translate legal obligations into a technical platform. The effort involves many stake holders like  legal experts, law enforcement agencies, software architects,  developers,  requirement analysts and security and privacy experts. They have to collaborate with a coherent goal to achieve data protection privacy by design \cite{PrivacyByDesign}. Work done by various stake holders in different areas  have been categorized in Table \ref{table:relatedork}. There is excessive antagonism between the development style of product in software industry and legal tenets of data protection framework. Several limitations and challenges exist in the legal framework that make implementation difficult.  A set of dichotomy between GDPR standards and system design perspective  has been discussed in \cite{7sin}. GDPR obligations like data storage, data deletion, data reuse are  challenges in the real world. Gruschka \emph{et al.} \cite{BigDataPbD} have discussed about data protection challenges while processing big data. They have suggested to use of  anonymity and de-identification techniques while processing big data. Esteve \cite{BuisnessOfPersonalData} has discussed the use of personal data by Google and Facebook for advertisement and business purposes and how this will affect protection of personal data under GDPR. Fuller \cite{DigitalPrivacyAFailure} argues why privacy is a failure till now. The debate is whether a company should collect data freely to provide a service, or they should impose a fee for the service, in turn  to protect the privacy. Similarly, \cite{Onpurpose} has analyzed GDPR challenges and limitations with respect to whether it is for a purpose or it was a necessity. 

\par 

\definecolor{LightCyan}{rgb}{0.88,1,1}
\definecolor{Gray}{gray}{0.9}
\begin{table}[htbp]
 \begin{center}
 
   \begin{tabularx}{\textwidth}{|X|X|}
           \hline
           \rowcolor{LightCyan}
				\textbf{Area} &  \textbf{References} \\
			\hline 
			Challenges and limitation in achieving goal of Data Protection & \cite{7sin}, \cite{BigDataPbD}, \cite{Onpurpose}, \cite{BuisnessOfPersonalData}, \cite{DigitalPrivacyAFailure}, \cite{Necessasry_IDPL} \\
		   \hline
		   Architecture & \cite{Architecture_DPbD},  \cite{Architecture_SoftwareIndustry}\\
		   \hline
		    Data Protection Properties & \cite{BlockchainMutability}, \cite{EfficacyOfGDPR}, \cite{FiveYear_RTF}, \cite{UninformedConsent} \\
		    \hline
		    Review of implementation of GDPR Policy in the system & \cite{TakeSomeCookies}, \cite{DPbD}, \cite{BigDataPbD}, \cite{BeforeAndAfterGDPR} \\ 
		   \hline
      \end{tabularx}\caption{Related work in the area of personal data protection}  \label{table:relatedork}
	%\begin{tabular}{|l|l|l|l|l|}
	%\end{tabular}
  \end{center}
\end{table}

After the implementation  of data protection framework, each data fiduciary has to change their system according to legal requirements of the framework. Hjerppe \emph{et. al} \cite{Architecture_SoftwareIndustry} and \cite{Architecture_DPbD} have analyzed and proposed software development models and threat models that can  be implemented as per the obligations of GDPR.  The other kind of works are based on procedure to accomplish the tenets of the framework. For instance \cite{EfficacyOfGDPR} has examined the importance of  ``right to forgotten'' covenant of GDPR. Similarly, \cite{BlockchainMutability}  has discussed about deletion of data stored in the blockchain from the perspective to ``right to forget''  of GDPR. Many authors have analyzed efficacy of present system: Such as, \cite{TakeSomeCookies} has examined  the privacy policy of cookies  by various websites after the implementation of GDPR. \cite{DPbD} has described the use case of cyber trust project for security of smart home environment. They have shown that, processing of personal data in cyber trust project  follows the policy of GDPR. Likewise, \cite{BigDataPbD} has analyzed  use case of two project ``SWAN''  and ``OSLO''. In this it is  investigated that, policy of GDPR is implemented effectively for the processing of big data and proper anonymization and de-identification has been followed as per the policy of GDPR.  

\par  Our work is completely different from all the previous work. To the best of our knowledge it is the first paper assessing the effectiveness  of PDPB. We have  described the technical challenges for each major obligation of PDPB framework and provided a set of properties and technical methods to help translate the legal covenant  into technical standards. We have described set of cryptographic methods that may help to implement PDPB's obligation. We have argued  how proper implementation of cryptographic   methods can provide better privacy, data protection and effective control on personal data.

\section{Background}
In this section we first discuss the  Personal Data Protection Bill and then give some definitions of the cryptographic primitives that we will propose to use. 

 \subsection{Personal Data Protection Bill (PDPB): 
 {\emph{What it is and how it will impact data processing?}}}
 In 2017,  \emph{right to privacy} was declared as fundamental right under the constitution of India \cite{RightToPrivacy}. The judgment by the Supreme Court of India, declares: 
 \emph{``the right to privacy is protected as an intrinsic part of the right to life and personal liberty under Article 21 and as a part of the freedoms guaranteed by part III of the constitution''}.
 
 In this digital era, no one is trustworthy over the Internet.  The principal concern  is to determine when an individual or organisation can share its personal data with others for processing and how the privacy of the personal data  will be protected. These concerns  demand an urgent requirement to develop a legal framework to protect privacy of individuals, to  protect the data of individuals  and to prevent against data breaches. In order to ensure this, Central Government of India established a committee lead by justice B. N. Srikrishna to study the existing  challenges and to establish a legal data protection framework to address the data privacy of individuals. The objective was  \emph{``to ensure the growth of the digital economy while keeping personal data of citizens secure and protected''}. The committee has proposed a  data protection framework known as the \emph{ Personal Data Protection Bill-2019 (PDPB)} \cite{IDPL}  which is the first step towards India's data privacy journey.
 
 \subsubsection{Basic constituents of PDPB}
 
\begin{figure}[!ht]
     
     \includegraphics[width=\textwidth]{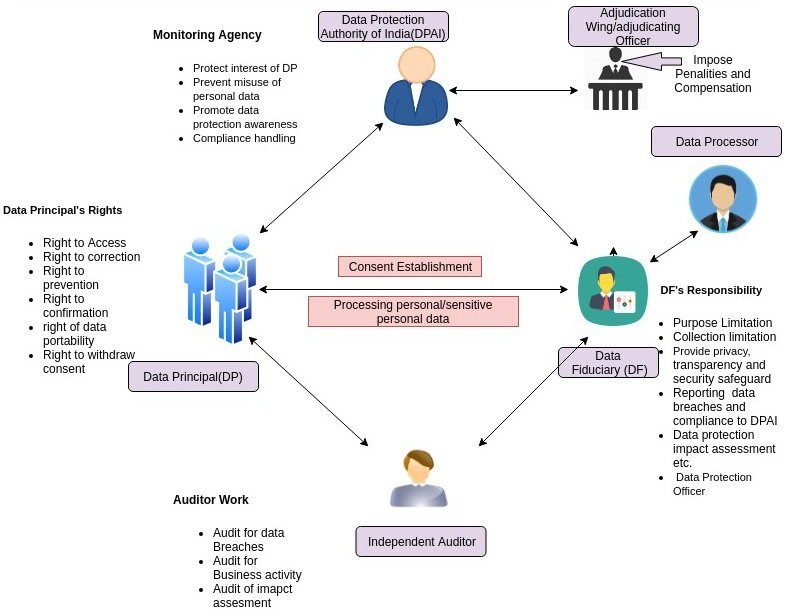}\caption{Primary entities of PDPB }\label{fig:idpl_framework}
\end{figure}

 Figure \ref{fig:idpl_framework} shows the primary entities of PDPB framework. 
 \par 
 \textbf{Data} is any information, opinion, facts, concepts and it can be categorized as health data, biometric data, genetic data, financial data etc.  \textbf{Personal data} is the  \emph{``data about or relating to a natural person who is directly or indirectly identifiable, having regard to any characteristic, trait, attribute or any other feature of the identity of such natural person, or any combination of such features, or any combination of such features with any other information''}. This bill also emphasises on some special types of data called as \textbf{sensitive personal data} that requires more   security and safeguard. It includes \emph{health data, financial data, sex life, sexual orientation, biometric data, genetic status,  caste or tribe, political or religious belief or affiliation}. Another category of data is \textbf{\emph{critical personal data}} which requires higher degree of protection and would be processed within India only. The data belonging to this category is not yet defined but in future the complete list will be notified by the Central government of India. 
 
 The entities involved are as follows: 
The owner of the data is known as \textbf{Data principal}. 
 \textbf{Data Fiduciary} is defined as \emph{
``any person, including the state, a company, any juristic entity or any individual who alone or in conjunction with others determines the purpose and means of the processing of personal data"}. 
\textbf{Data processor}, who either can be a data fiduciary or a third party who processes the data on the behalf of data fiduciary. This bill explicitly mentions  that the responsibility of data fiduciary is to protect the data of an individual. 

Other entities involved could be independent auditors who perform data audits, the Data Protection Authority of India (DPAI) that sets guidelines and makes legal decisions based on the inputs of the data principal, data fiduciary and independent auditors.  
%It has to perform the overall activity such that it protects data principal's personal data.

\subsubsection{Brief description of obligations of  PDPB}
In Table \ref{table:articlewiseIDPL}, we have shown organization of PDPB framework.

%Articles 1-3 describes aim and scope of the framework and defines various definition  used in the bill. Next, article 4-8 establish grounds for processing of personal data. 
\par
Data fiduciary has to collect, disclosed, shared and processed personal data  for purposes that should be  \emph{clear, specific, reasonable and lawful} (as per Article 4). \emph{Clear} means, data principal knows the exact reason for which data has been collected. Further, data shall be processed only for  reasonable and \emph{specified purposes}  (as per Article 5). Such purpose will be identified and defined by the data fiduciary before the data collection and shall be disclosed to the data principal at the time of collection. Data fiduciary has to perform \emph{fair and reasonable processing} of data. In particular, any individual or organization who is processing personal data owes a duty to the users to process such personal data fairly and reasonably  that respects the privacy  of the data principal (as per Article 5(a)). Fair data processing will impact the technical design of a system and the way data is collected and processed. Therefore data fiduciary has to collect least and minimum data as possible according to the \emph{``collection limitation"} requirements (as per Article 6).  

\par
Before collection and processing, data fiduciary has to provide \emph{the necessary notice} (as per Article 7) to the data principal. The notice will contain few basic prerequisite such as purpose of collection, consent form, nature of data being collected, information about any cross border transfer etc. Data fiduciary also needs to maintain quality of personal data during processing. This means that  data must be accurate, complete and not misleading (as per Article 8).  

%Articles 12-24 are related to the processing of personal and sensitive personal data.\emph{The processing of the personal data} shall be done based on the \emph{consent of data principal} (as per Article 12). 
\par
\emph{Consent} is required before the commencement of data processing (as per Article 11). Consent would  be \emph{free, informed, specific and clear}. Data  fiduciary will have responsibility to keep a proof that consent has obtained from data principal before the processing. The protection framework puts more restriction on the processing of \emph{sensitive personal data}. The processing of such kind of data  will require \emph{explicit consent} (as per Article 11(3)) from the data principal. Explicit consent means  data fiduciary has to draw attention to the data principal about the sensitivity of personal data, the reason why such collection is necessary and possible consequences. Data principal will also have to acknowledge explicitly. A \emph{consent manager} will also established by data fiduciary which will record and manage all the consent obtained from data principal. The Data Protection Authority of India (DPAI) shall publish list of personal attributes which would be considered as sensitive personal data periodically (as per Article 15).  
%Personal and sensitive personal data can also be processed for the interest of the state.

\begin{table}[htbp]
		\begin{center}
			
			\begin{tabular}{|l|l|}
				\hline
				\textbf{Articles} &  \textbf{Details}  \\
				\hline
				Article 1-8 & Objective and  establishment of grounds for  processing of personal data \\
			     
			     \hline
				Article 9-10 & 
				Data retention policy and Accountability\\
				\hline
				Article 11 & Definition of consent and explicit consent\\
				
				\hline
			    Article 12-14 & Norms for processing of personal data without consent \\
			    \hline
			    Article 15 & Criteria for consideration of personal data as sensitive personal data  \\
			    
				\hline
				Article 16 & Grounds for processing of personal and sensitive personal data of children \\
				\hline
				 Article 17-21 & Establish the  rights of data principal  \\
				\hline
				 Article 22-24 & Privacy, transparency, consent manager and security safeguards 
				 \\
				 \hline
				  Article 25-32 & Significant data fiduciary, data breach, data audit, 
				  \\
				  &data impact assessment and data protection officer  \\
				\hline
				Article 33,34 & Grounds for transfer of personal data outside of India, critical personal data, \\
				\hline
				Article 35-40 & Mandate about processing of personal data for other purposes like \\
				& security of state, for law or legal processing, journalistic purpose, \\
				& research or statistical purpose etc. \\
				\hline
				Article 41-56	& Establishment, responsibilities and power of
				\\ & data protection authority of India\\
				\hline
					Article 57-85 & Covers penalties, liability, establishment of appellate tribunal and
					\\ & execution of other offences \\
				\hline
				Article 86-98	& Miscellaneous power of central government,\\
				& grounds for framing digital India policy and norms for processing biometric data   
			    \\
				\hline
				
			\end{tabular}
			\caption{Organization of articles of PDPB-2019 } \label{table:articlewiseIDPL} 
		\end{center}
	\end{table}
\par  
The \emph{processing of children's data} will be in such a manner which  protects the rights of children (as per Article 16). Data fiduciary has to verify the age of children and has to obtain consent from parent or guardian before processing the data.  DPAI will specify the procedure and the appropriate mechanism to conduct age verification under this regulation. These procedure may vary depends on the type, volume and sensitivity of personal data corresponds to the children. Consequently, data fiduciary has to implement it  technically. Further, based on the nature of commercial website, online services offered  and the volume of children personal data being processed, DPAI will categorize few data fiduciary as \emph{guardian data fiduciary}. They are not allowed for profiling, target based advertising or any other activities which may lead harm to the children.
 \par
 
  The data protection framework  provides various rights to the data principal such as  \emph{the right to confirmation and access} (as per Article 17), \emph{the right to correction and erasure} (as per Article 18), \emph{the right to data portability} (as per Article 19), and \emph{the right to be forgotten} (as per Article 20). Data principal may ask what data is being processed from data fiduciary. He has right to obtain the summary of processed personal data. Data principal may request to update - alter, correct or erase his personal data. He may also request  data fiduciary to prevent or restrict continuing disclosure of personal data. 
  
  %\textcolor{blue}{A question is how to the data fiduciary can prove that the update has been considered and data has been re-processed?} 
 
 \par
 The design of a system should provide privacy, transparency and security to the individual automatically. Therefore data fiduciary has to take necessary steps to implement \emph{privacy by design} (as per Article 22), \emph{transparency} (as per Article 23) and security safeguard (as per Article 24). Here transparency means disclosure of steps which has used to protect personal data and security safeguards are like the  use of suitable de-identification, encryption mechanism, integrity protection and access control mechanism.

  %Data fiduciary shall implement data processing activity in a transparent manner which preserves the intended rights of data principal effectively. Data fiduciary and data principal will implement appropriate  (as per Article 31). Data fiduciary has to prevent  misuse, unauthorized access, disclosure, modification or destruction of personal data.  
  \par
  Data fiduciary may conduct \emph{data protection impact assessment} before commencement of processing of personal data (as per Article 27). Impact assessment depends on nature, scope, context and volume of personal data. Each data fiduciary will maintain up-to-date \emph{records}. Records may contain activities such is important operation, impact assessment, periodic review etc (as per Article 28). Also, an \emph{audit} may be conducted periodically by an independent auditor approved by  data protection authority to review data processing policy, to analyze risk and to verify data protection impact assessment (as per Article 29). Each data fiduciary shall assign a \emph{data protection officer} (as per Article 30) who will advice the data fiduciary for matters related to data processing.  The data protection officer also works as a point of contact for the data principal as well as for the data protection authority. The fiduciary who has not established their office in India and performing processing activities of personal data has to assign a data protection officer within India itself. All \emph{data breaches} (as per Article 27) which may harm the data principal, shall be reported to DPAI by the data fiduciary.

  %\emph{Data protection authority of India (DPAI)} (as per Article 49)  shall be established by the government of India which will monitor the data protection activity.  
  %The report will contain nature of personal data, number of affected data principal, possible consequences of data breaches and possible measures taken by data fiduciary. DPAI will decide further, whether this breaches should be reported to data principal or not depends on sensitivity nature and severity of harm of personal data harm.The data fiduciary shall conduct  an annual \emph{data audit} and  processing of personal data activity by an independent data auditor (as per Article 35).
  \par
  PDPB puts multiple restrictions on \emph{cross-border transfer of personal data} (as per Article 33). Sensitive personal data can be transferred outside India but data fiduciary has to store at least one copy  within India. The DPAI will determine which kind of sensitive personal data can be transferred. It will based on nature of data, international relation of India with the countries and  international agreement.  Explicit consent of data principal is required before such transfer. \emph{Critical personal data} shall be processed within India itself. 
  
  \par
  Based on nature, volume and severity of data being processed by a data fiduciary or class of data fiduciary can be called as \emph{significant data fiduciary} (as per Article 26). Each social media data fiduciary which is intimated as significant data fiduciary will verify the account of all user who uses their services in India or use their service from India (as per Article 28).

 \subsection{Basic cryptographic primitives}
 Cryptography is a practice of of making secure communication between sender and receiver in the presence of adversaries. Cryptographic algorithms  design  can provide security, integrity and privacy of data. Here we discuss recent techniques of  cryptography that can be used in data protection. 
 
 \par
 \textbf{\textit{Encryption}} \cite{mc}
is a process of converting message (plaintext) to ciphertext, in such a way that the ciphertext does not reveal any meaningful information about the  message. A cryptograhic key (a random string) and message is supplied into a well defined encryption algorithm to obtain ciphertext. The receiver who obtains  ciphertext and has knowledge of key can recover the plain text using a decryption algorithm. Mainly there are two kind of encryption symmetric and asymmetric.  Ciphertext provides confidentiality or privacy and security while data is moving or at the rest. An adversary observing message and ciphertext pairs cannot distinguish between ciphertexts of two different messages. The limitation of encryption is, privacy and security can not be guaranteed when sensitive data is disclosed improperly.      
 
 \par
 \textbf{\textit{Digital signature}} \cite{mc}
 is  a cryptographic scheme which verifies the authenticity of sender and integrity of a digital document (message).  Sender creates a signature on the message and sends to the receiver. Upon receiving both message and signature  receiver can verify that message has not been altered and is signed and send by the  claimed sender itself. Signatures should be unforgeable, such that an adversary who observes message signature pairs cannot construct a valid message, signature pair. Data fiduciary can use signatures to prove authenticity, integrity and non-repudiation.  
 %Its application area could be  financial transactions, email communication, software distribution, websites etc. 

\par
\textbf{\textit{Message authentication}} \cite{mc}
message authentication is scheme which proves that message is not altered (data integrity) while in transit and receiver can verify the source of the message.   Message authentication can be implemented through \emph{ message authentication code(MAC) or Authenticated Encryption(AE)} schemes. Along with message a tag value is send to the receiver which can be used to verify authenticity and integrity of the message. While signatures can be verified by any party with public information, MAC and AE need secret keys (information) for verification. 

\par
\textbf{\textit{Zero knowledge proofs}} \cite{ZKP}
Zero knowledge proofs (ZKP) are useful when a party wants to prove that it possesses some information, without revealing the information. For example, proving that a person is above 18 years without revealing the age (or showing the identity document).   Generally, ZKP has two parties prover and verifier. Prover(such as data principal, owner of secret information) can prove to the data fiduciary that he knows a personal data ($x$) without disclosing any information apart from fact the he knows a value $x$. 

\par
\textbf{\textit{Secure multiparty computation}} \cite{SMPC}
Secure multiparty computation (SMPC) allow multiple parties to perform computation on encrypted or private data without revealing the data to the other parties. SMPC  gives more privacy over user's private data. Two parties( such as two hospitals) can compute a joint function (say total number of cancer patients) jointly while keeping their inputs private.

\par
\textbf{\textit{Homomorphic encryption}} \cite{hmomorphic}
Homomorphic encryption allows computing on encrypted data. 
Data principal shares personal information to data fiduciary possibly over encrypted channel. These information are decrypted and used by data fiduciary and sometimes can be misused. To prevent improper disclosure, homomorphic encryption(HE) allows user to share their personal data without loosing confidentiality and privacy.

\par
\textbf{\textit{Searchable Encryption}} \cite{Searchable_Encryption}
Searchable encryption is a cryptographic technique that can query on encrypted data stored in outsourced servers. The search neither leaks the response (search results) to the server, nor does is the server aware of the query itself. Popular queries include single/conjunctive keyword search, range queries etc. 
%. In this scheme, along with cipher text of the document users also stores a tag value which are mainly the keywords associated with the document. Tag  helps to find  relevant  document by service provider  when search document is queried from the users.  It will enable users to search and retrieve the document from the semi trusted cloud server without decrypting and viewing contents of the document.

\par
\textbf{\textit{Blockchain}}
Blockchain \cite{Blitcoin} is a distributed, tamper proof, immutable and decentralized ledger. The data recorded in the blockchain are publicly verifiable. All transactions originating from users are propagated to the network and validated by multiple validators. This improves the trust of the system. Immutability guarantees non-repudiation, meaning that users cannot deny making the transactions. Blockchains guarantee that all transactions are recorded, verified and cannot be retracted.  In the literature framework has been proposed to integrate blockchain for the data protection \cite{BC36_blockchainPDP}. 
\par
\textbf{\textit{Differential Privacy}} \cite{differential_privacy} It allows organization to share aggregate information of their users while protecting private information. Differential privacy mathematically ensure that any one obtaining the analysis result would make the same conclusion about individual's private information whether or not his data is included in the input for the analysis.

\section{Comparison of GDPR-EU and PDPB}

 After the implementation of GDPR in May 2018 \cite{GDPR}, many organizations that are providing their service in the European Union have changed their  data processing policy as per GDPR standards \cite{TakeSomeCookies}. PDPB framework is proposed with respect to the Indian context and has been inspired by GDPR. 
 Both protect individual's in the processing and free movement of personal data. 
 %In the context of GDPR,  the  data principal is called ``data subject'' and data fiduciary ``data controller''. 
 %But both are just different naming convention. 
 There is a good deal of similarity between GDPR and PDPB with few differences.
 We have analyzed the major differences that exist between GDPR and PDPB, why such differences exist, what are the effects and whether these differences makes PDPB stronger or weaker. Table \ref{table:compare} compares GDPR and PDPB.
%write about strong point where GDPR bill was strongly written but indian bill has lag. for example right to prevent disclosure, right to prevent automatic decesion making processing. \par

 \par
The primary difference is that both frameworks have different approach for  \emph{sensitive personal data and critical personal data} processing and its storage. In GDPR, European Union allows \textbf{\emph{cross border transfer of data}} to non-EU countries/organisations if commission decides that such transfer has an adequate level of protection. The adequate level of protection includes criteria such as nature of the data, law and enforcement in non-EU country/territory, international relations  and the commitment of data fiduciary for the security and safeguard of data. Such transfer will not require any  specific authorisation.  PDPB enforces  more restrictions on cross boundary transfer of data. First of all, the data protection authority of India will specify the definition of sensitive personal data and critical personal data. For the case of sensitive personal data, DPAI will authorize a party to be transferred to outside of the border.  DPAI  will also authorize whether such transfer has an adequate level of protection.  Additionally, data fiduciary has to notify data principal explicitly regarding cross-border of transfer. Another constraint of PDPB is, data fiduciary has to maintain a  local copy  because Indian government is more concern about availability of data. Critical personal data has to be be stored and processed only within India.  Definition of critical personal data is yet to be disclosed but it could be Aadhar data (unique identification number), PAN data (permanent account number, issued by income tax department), digi-locker data. Presence of such control might be because the  government is concerned about the security of critical personal data after data breaches were reported \cite{breach_aadhar}.

%May be due to all these concern  localization of critical data  has been kept. 
 % Contents in the second and third columns will be auto-wrapped,
% so that the entire table will fit the text width nicely
\definecolor{LightCyan}{rgb}{0.88,1,1}
\definecolor{Gray}{gray}{0.9}
\begin{table}[htbp]
 \begin{center}
   \begin{tabularx}{\textwidth}{|X|X|X|}
           \hline
           \rowcolor{LightCyan}
			\textbf{Detail} &  \textbf{GDPR-EU} &  \textbf{PDPB} \\
			
			\hline 
			Status & Implemented & Draft \\
			\hline
            The entity whose data is being processed & Data subject & Data principal \\
			\hline 
			The entity who is processing data & Data controller
				 & Data fiduciary\\
			\hline
			Cross border transfer of sensitive personal data & May be transferred without any restriction, if commission/authority assure appropriate safeguard and protection & May be transferred if data fiduciary follows the guidelines of authority.
			Also, data fiduciary has to store one local copy within India. \\
			\hline
			Critical personal data & May be transferred without any restriction, if commission/authority assure appropriate safeguard and protection & Processed and stored within India\\
			\hline
			Data breaches & Data controller may inform to data subject, in case of high risk   & Data protection authority  will  decide whether it should be reported to data principal or not\\
			\hline
			Right to object & Data subject may object further processing of data & Not specified\\
			\hline
			Right to restriction & Data subject may restrict processing & Not specified \\
			
			\hline
			Children consent & Approval of parental authority is required to process children personal data for children below 16 years & Children data shall be processed with appropriate safeguard which protects the rights of children; Age verification and parent's/guardian's consent is required\\
			
			\hline
			Children data processing & Not specified & Guardian data fiduciary shall be barred from profiling, tracking, behavioural analysis and target based advertisement that can cause significant harm to the child. \\
			\hline 
			Right to object automated decision making & Data subject may have right not to participate in automated decision making like profiling which can harm  him/her significantly
				 &  Not specified\\
			\hline
			Right to access & Data subject may obtain copy of personal data being processed  & Data principal May obtain summary of personal data being processed \\
			\hline
			Social media significant data fiduciary & Not specified  & Specified, have to maintain user verification records also \\
	
			\hline
			Data audit & \checkmark & \checkmark\\
			\hline
			Consent Manager & Not specified & Record and manage consent of data principal\\
			\hline
			Trust score & \xmark & \checkmark\\
			\hline
			Certificate & \checkmark & \xmark\\
			\hline
			
            \hline
      \end{tabularx}\caption{
      Comparison of GDPR-EU and PDPB
      }\label{table:compare}
	%\begin{tabular}{|l|l|l|l|l|}
	%\end{tabular}
  \end{center}
\end{table}	

%Code of conduct & \checkmark & \checkmark\\
			%\hline
			%Data protection officer & \checkmark %& \checkmark\\	
 
 \par
 Another major difference is, GDPR and PDPB have different approach to report \textbf{\emph{personal data breaches}} to the data principal. 
In PDPB, data fiduciary will report data breaches directly to the data protection authority. Later, taking into account the severity of harm and the necessary action will be taken. The  authority may decide  whether it should be reported to the data principal.  In GDPR, data subject has more power in the case of data breach. According to this, data controller has to  report data breach directly to the data principal. Subsequently if data fiduciary finds that data breach may result in high risk to the rights and freedom  of natural person  such breach shall be reported with undue delay. This reporting  will also disclose the necessary steps and remedial action which have been  performed by the data controller to mitigate the risk. Thus, PDPB regulation provides more power to the DPAI instead of data principal in the case of breach reporting. 

%If data fiduciary is strong he can influence and influence of data fiduciary, central government or if authority become malicious then breach can be easily concealed. 

%\textbf{Processing of sensitive data and special category data} Both in PDPB and GDPR, sensitive personal data may be processed after the explicit consent of  data subject. Data fiduciary has to inform explicitly about the processing of sensitive personal data and its possible consequences.

\par
Further, PDPB introduce the  definition of \textbf{\emph{significant data fiduciary}} which is not present in the GDPR. Based on the volume of personal data processed, sensitivity of personal data, risk of harm in processing or any other criteria DPAI will categorize some data fiduciary as  significant data fiduciary. They have to register themselves with  the DPAI. Authority will direct and monitor him directly regarding  the implementation and important operations such as data audit, data protection impact assessment, maintenance of record, establishment of data protection officer or direction for security and safeguard of the data. Similarly,  authority will further classify and publish a list of \textbf{\emph{social media significant data fiduciary}}. These will be from the area who provides online social media services.  Such data fiduciary would have an  additional responsibility of  identification and verification of every users who are getting service from or within India. 
PDPB also introduced the concept of  \textbf{\emph{consent manager}}. Consent manager is an entity who will manage and update all the records of consent provided by data principal. 
%Such an entity is not present explicitly  in GDPR but can be easily derived from the definition of consent.   

\par
 Both frameworks are concerned about an adequate protection during the \textbf{\emph{processing of children personal data}}.
 As per the PDPB, a person below 18 year will considered child in India. PDPB mandates age verification prior to data processing, in addition to consent from the authorized guardian. 
 In GDPR, children (below 16 years) data  can be processed only if the holder of parental responsibility gives consent.  
 %In PDPB both age verification and consent of parent or guardian are required for processing children's (below 18 years) data. %Each data fiduciary has to set up an extra security and safeguard that can protect rights of the children during data processing.
 PDPB will specify appropriate mechanism that can be used for age verification. PDPB has introduced the concept of  guardian data fiduciary. They are not allowed to do profiling, targetted advertising or any action which can harm the children. Such restriction is not specified in GDPR. Overall, for children data processing, PDPB is stricter than GDPR. 
 \par 
 Taking into account of rights of data principal, both framework has different definitions and approach. GDPR gives more rights to data subject. 
 \par 
 Data subject has right to get  information in both framework when  \textbf{ data is not collected directly from data subject}. In PDPB, if data fiduciary has not collected personal data directly from data principal then it has to disclose source of such data before the commencement of processing or as early as reasonably practical. On the other side in GDPR, if personal data has not obtained from data subject then data controller has to provide the source the collection,  purpose for which data will be used, categories of personal data obtained, further recipient of this data and any other relevant  information.
 Further, if data controller does \emph{profiling, automated decision making or any behavioral analysis} then data subject may request information regarding this under GDPR. If it is not possible to give all such details then the detail of the idea behind the logic involved, significance of processing and possible consequences may be provided. Such clause is  not present in PDPB. 
 
\par 
Under GDPR, if any data controller wishes to use \textbf{\emph{personal data for further processing}} other than the purpose for which it is collected, the controller has to inform to the data subject about the purpose of processing along with any other relevant information. PDPB does not have any clause regarding such further processing. Similarly, In GDPR  
data subject solely have right not to participate in \textbf{\emph{profiling and  automated decision making}} which might harm him  significantly. In PDPB, data principal does not have such rights.

%This objection shall not be applicable if such criteria is necessary to enter into the contract of processing or such processing is done by union member or member state or it is based on explicit consent of data subject.

\par 
 GDPR has  \textbf{\emph{right of restrictions}} and \textbf{\emph{right of object}} for data principal. Data principal may restrict his processing instead of deletion. Restriction can be imposed on the presence of conditions such as the  accuracy of personal data processed by controller is disputed or data principal believes that processing is unlawful. Restriction can also be requested if controller can not delete data. It can be the situation where the storage of personal data is  necessary for the establishment, exercise  or defence of legal claim.  Similarly, at any time data subject may object the processing under GDPR.  Instead of right to restriction, PDPB  has \textbf{\emph{right to forgotten}}. Using this  data principal can restrict or prevent continuing disclosure of his personal data. But it is applicable  only by the order of an adjudicating officer.

%\par 
%\textbf{Right to erasure} As per Article 17 of GDPR,  data subject has  right to erasure of personal data. Data subject shall have right to ask data controller to erase the personal data without undue delay if certain rules applies like, \emph{``the personal data are no longer necessary in relation to the purposes for which they were collected or otherwise processed''}  or data subject has withdraw his consent for further processing. Right of erasure does not exist in PDPB. Instead of this, Indian data protection framework allow data principal to obtain right to forget. In this data principal  can request to the data fiduciary to restrict or prevent disclosure of personal data to the third party or from further processing (under the Article 27.

%\textbf{Processing of personal data which not require identification}
%If the purposes for which a controller processes personal data do not or do no longer require the identification of a data subject by the controller, the controller shall not be obliged to maintain, acquire or process additional information in order to identify the data subject for the sole purpose of complying with this Regulation(copy).
\par
Some minor similarity and differences also exist which are mentioned in Table \ref{table:compare}. For instance, in  \textbf{right to access}
 data subject may obtain a copy of personal data undergoing in processing  under GDPR. While in PDPB, only a brief summary of personal data  may obtained. Similarly, PDPB has concept of trust score (a kind of rating on a scale)  while GDPR has concept of  certificate to the data fiduciary.

 \section{Major PDPB Regulations From System Design Perspective}
 To comply with the privacy regulations, data fiduciaries need to modify their systems and data processing procedures. 
 \begin{figure}[!ht]
     
     \includegraphics[width=\textwidth]{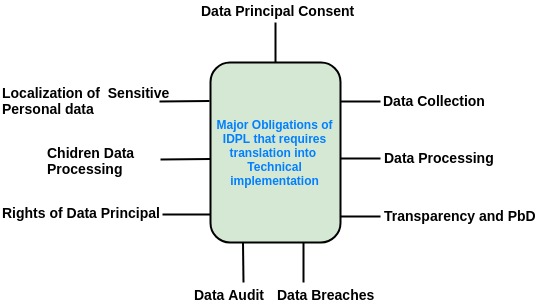}\caption{Major obligations that requires translation into technical implementation} \label{fig:obligations}
\end{figure}
  
  %The set of obligations that need to considered first and foremost have summarized in the figure \ref{figure:majorpoint}.
  To achieve the goal of obligations, the first and foremost requirement is to understand and interpret the legal requirements precisely. Data fiduciary needs to investigate what are the necessary data protection, security and privacy by design requirements.  Data protection goal can not be accomplished  until and unless the data fiduciary understands the explicit meaning of obligations. 
  %Figure \ref{fig:obligations} shows the various requirements of the bill. 
  
  Next step is to understand how the requirements can be achieved using privacy tools.  There are many tools and techniques  in the literature. We strongly argue that if such tools are used appropriately then it will help to  implement the law effectively.  This will help in achieving the goals of data protection bill. 
  
  For instance, the bill gives excessive power to the Government (as per Article 91). Central Government can ask for any personal data to design policy framework for the country or to promote digital economy. Data fiduciary has to reveal data to the Government. In such cases, we assert that  there are cryptographic tools and techniques which would not only help to implement data sharing but also preserve the privacy of data principal. For example, zero-knowledge proofs and  secure multi party computations can be used for data processing  without revealing the data \cite{ZPGS19,MZ17}.
  %For example ABC4Trust for privacy preserving credential management has already been deployed.  
  Another scenario where implementation of cryptographic techniques will be useful is that in sharing data about Aadhar number (unique identification number for Indian citizens) \cite{SP-Aadhaar}.  Individuals need to share biometric information to get government beneficiaries services. Zero knowledge proofs and fuzzy attribute based encryption \cite{SW05} can be used to verify individuals identity without revealing biometric information to the data fiduciary.       

 We have  considered only the major obligations (that we call verticals) from the framework. These are  shown in Figure \ref{fig:obligations}.    Analysis of each vertical is done primarily in two parts. First,  clauses in the  vertical are well identified and the security and   privacy  requirements are derived. Second, a description of appropriate tools and techniques has been provided against properties derived from the previous step. Wherever possible, an 
  explanation has also been provided regarding  why such properties are necessary. 
  
  % So,
  %We have provided only the high level overview of methods and we have proposed that it  will work as backbone for achieving all the goal of PDPB. 
   
  % Contents in the second and third columns will be auto-wrapped,
% so that the entire table will fit the text width nicely

\definecolor{LightCyan}{rgb}{0.88,1,1}
\definecolor{Gray}{gray}{0.9}
\begin{table}[htbp]
 \begin{center}
   \begin{tabularx}{\textwidth}{|l|X|X|l|}
           \hline
           \rowcolor{LightCyan}
				\textbf{S.No.} &  \textbf{Required Properties} & \textbf{Technological methods} & \textbf{References} \\
			\hline
			1. & Data consent & Digital signature, encryption and blockchain & Section \ref{subsection:proofofconsent} \\
			\hline
			2. & Limited data collection  & Data minimization techniques & Section \ref{section: datacollection}\\
			\hline
			3. & Data storage limitation  & Required expiry of data & Section \ref{subsection:storagelimitation}\\
			\hline
			4. & Privacy by design(PbD)  & Storage of encrypted data, searchable encryption, zero knowledge, secure multiparty computation, homomorphic encryption, anonymous and confidential storage &  Section \ref{subsection:pivacybydesign}\\
			\hline
			5. & Integrity protection and authentication & Authenticated encryption, continuous data authentication scheme, 2 way authentication, 2-factor authentication, multiparty authentication and time based authentication & Section \ref{section:dataprocessing}\\
			\hline
			6. & Security and safeguard & Encryption, authentication and authorization of data, client side encryption of cloud data, searchable encryption, secure multiparty computation, secret sharing & Section \ref{section:dataprocessing} \\
			\hline
			7. & Anonymization and de-identification of data & K-anonymity, l-diversity, t-closeness, differential privacy, tokenization, federation, hashing, binning  format preserving encryption & Section \ref{section:dataprocessing} \\
			\hline
			8. &  Access control &  Role based access control, access control list, policy based access control, geolocation based access control, attribute based access control, lattice based access control& Section \ref{section:dataprocessing}\\
			\hline
			9. &  Right to access, Right to correction & Blockchain base method & Section \ref{subsection:userright} \\
			\hline
			10. &  Right to forget & Erasure based proof & Section \ref{subsection:userright} \\
			\hline
			11. &  Data audit & Proof of storage, proof of consent, proof of possession, proof of retrievability, proof of sharing of data & Section \ref{subsection: audit} and Section \ref{subsection:blockchaindataaudit} \\
			\hline
			12. & Data breach & Data audit & Section \ref{subsection:databreach}  \\
			\hline
			13. &  Children data processing & Anonymous user credential, Parental control, zero-knowledge based proof&  Section \ref{subsection:children} \\
			\hline
			14. &  Location of sensitive personal data & Proof of location, geo fencing, proof of geo-location of data& Section \ref{subsection:localisation}  \\
		\hline
      \end{tabularx} \caption{Tenets of PDPB and various methods to implement it technologically } \label{table: tech_imp}
    \end{center}
\end{table}
  
 The essence of implementation  is  the translation of explored properties into technical implementation. To achieve this, a   summary of the list of vertical's and their corresponding set of tools has been shown in Table \ref{table: tech_imp}. Many cryptographic methods can be used to solve  security, privacy and data protection requirements of obligations.  In many cases we have shown that appropriate use of cryptographic solutions such as authentication, encryption, integrity, proof of storage, proof of possession, proof of location, zero knowledge proof, secure  multiparty computation and other techniques can provide a solid support for implementation of the data protection bill. A detailed discussion of each verticals is done next section.

 %In below sub section for each verticals a detailed description of interpretation is provided. Additionally, a guidelines is given how effectively such cryptographic tools and  algorithms can  used in order to achieve strong technological standard.
 
 \subsection{Data Principal Consent} \label{dataconsent}
 
% \textbf{Consent} come into consideration effect when the user agrees upon something \cite{consent}.
%\textbf{Consent} of users is required because,
As per the right to privacy, \emph{the user has right to set their choices, preferences, permissions, to decide access control on their personal information, to select with whom, where and how much personal information shall shared}\cite{consent}. To enforce this data protection framework specify the definition of user's consent. It states that consent is necessary and should be obtained before collecting and  processing of personal data. Many organizations are collecting user's personal information for their monetary purpose, social and personal benefit, business analysis, target-based  advertising or  for behavioural analysis \cite{cambridge, DigitalPrivacyAFailure, BuisnessOfPersonalData}, which may affect users privacy. Therefore, consent as a legal evidence will help to develop trust  and prevent data fiduciary and third parties from data misuse. 
 
 %It enforces that  data collected for one purpose cannot be used for another purpose without  owner's permission.
 %The processing of personal data shall be done only after \emph{valid and informed consent } from the data principal.In particular, collecting and sharing of personal data  should be done in the presence of a user's knowledge.
 \par
 
  PDPB states the nature and scope of the consent for data processing. According to Article 11(1) of PDPB, \emph{``data may be processed based on consent.."}. Data fiduciary must notify the data principal regarding the data collection and should obtain the consent. This is known as \emph{informed consent}. Further, Article 11.2(a) - 11.2(d)) states: \emph{``consent shall be free, informed, specific, clear.."}. Data principal also has the \emph{right to withdraw their consent} (as per Article 11.2(e)). Subsequently, for sensitive personal data, \emph{explicit consent} is required (as per Article 11(3)). Explicit consent means the consent is \emph{informed, fair, specific and unambiguous}. In other words, data principal has to acknowledge before such data collection. This acknowledgement can be done  by a statement, an affirmation, an action, selection, choices or any other kind of agreement. Next, data Fiduciary as per Article 11 (5), \emph{``shall bear the burden of proof to establish that consent is given by data principal''} so that he can prove transparency to anyone in the processing .   Finally, data fiduciary has to implement an entity \emph{consent manager}  through which he can record, manage and process all of the consent (as per Article 23(3)).

 \par
  We will now discuss set of properties that need to be solved to implement the consent technically. Further, we have discussed whether or not current model of consent exhibits such properties. Later we have discussed  the high-level  idea of how it can be implemented.    
  
 \subsubsection{Proof of Consent} \label{subsection:proofofconsent}
  \label{subsec:proofofconsent}
The Data Protection Bill presents consent from legal perspective. This essentially means that parties are questionable in the court of law, if they deviate from the terms and conditions in the consent form. Legal action is taken after an event and requires evidences. These evidences can be (possibly) encoded within the software to ensure that the consent properties are implemented accurately and transparently. We call such encoded evidence or proofs as  \emph{proof of consent (PoC)}. These proofs will prevent data fiduciary from deviating from the legal contract (as in the consent form). It will help to resolve  disputes if they arise, or it can be presented to the jurisdiction of DPAI for any legal investigation. Such PoC should be verifiable during data audit in order to detect malicious activities. At a very high level, the following properties are required to be implemented  to construct proof of consent:
 
 \begin{enumerate}
     \item Data fiduciary should establish a free, informed,  specific, and clear method to implement consent. Consent can be established through opt-in methods so that user will have an option to click check boxes and select his choices. Data fiduciary should maintain purpose explicitly for which consent is being made. A separate consent can be established for different purposes. 
     \item Data fiduciary should implement a proof that a consent is established with the data principal.  If data principal claims that data fiduciary is processing data without their consent, then data fiduciary can prove legally that consent was indeed established.
     
    \item As per the bill,  data fiduciary should mention the list of third parties with whom data will be shared in consent form. List of third parties will help data principal to keep track of personal data. But as of now, data fiduciary does not keeps transparency in disclosure of third party. Such transparency is required to implement technologically.
    
    %(NOT CLEAR)
    
     \item Data fiduciary should implement a  proof that personal data has been shared with those third parties only, for which data principal has given consent.  
     
     \item Data principal has right to withdraw consent. Data fiduciary has to provide a method using which  consent can be withdrawn. Data fiduciary should keep a proof of withdrawal of consent. Also, data fiduciary  should  maintain a proof that all the third parties have been intimated about such withdrawal and both have stopped processing and sharing this data.  Such proofs are required because as per the law, no data  processing should be done after consent has been withdrawn. Violation of this may lead to punishment/penalty. On the contrary, this violation will work only if it came under the observation.  Otherwise behind the wall data fiduciary may use this data for monetary purpose even after consent has been withdrawn. To the best of our knowledge, this problem is rather difficult to solve without legal interventions.  

     \item Data fiduciary should maintain a record of whom, when, how and why consent has been established with data principal.
     
     \item Data fiduciary should allow data principal to change his choices or preferences. If such changes have been approved by data principal then further data processing will be done as per the  new consent. Data fiduciary should maintain a proof that modified consent is recorded and processing is being done on modified consent.
     
     \item In the case of explicit consent, data fiduciary shall make a clear and concise statement. Data fiduciary will write statement about sensitivity of data, why it is required and  will ask for agreement of data principal explicitly before processing it. He/she should keep a proof that explicit consent has been obtained before processing of sensitive personal data.
     
     \item Data fiduciary may review the consent policy time to time and should inform the data principal accordingly. He can update the list of changes in the consent established previously. Data fiduciary has to provide an option to the users either they may change their preferences as per the new consent policy or they may withdraw their consent  from further processing. In both cases, data fiduciary should keep a proof of such modifications.
     
     \item Data fiduciary has to follow the guidelines of PDPB. If any change in the bill occurs regarding consent then data fiduciary has to review it and have to modify their policy as per the new standards.   He will notify the same to the users and ask for necessary  modification of  consent before further processing. Data fiduciary should keep a proof of such modifications.
     
     \item Data fiduciary should maintain a method of proof that he is not collecting any personal data automatically without user's consent.
     
\end{enumerate}

 \par
\begin{figure}[!ht]
     \includegraphics[width=\textwidth, height=5cm]{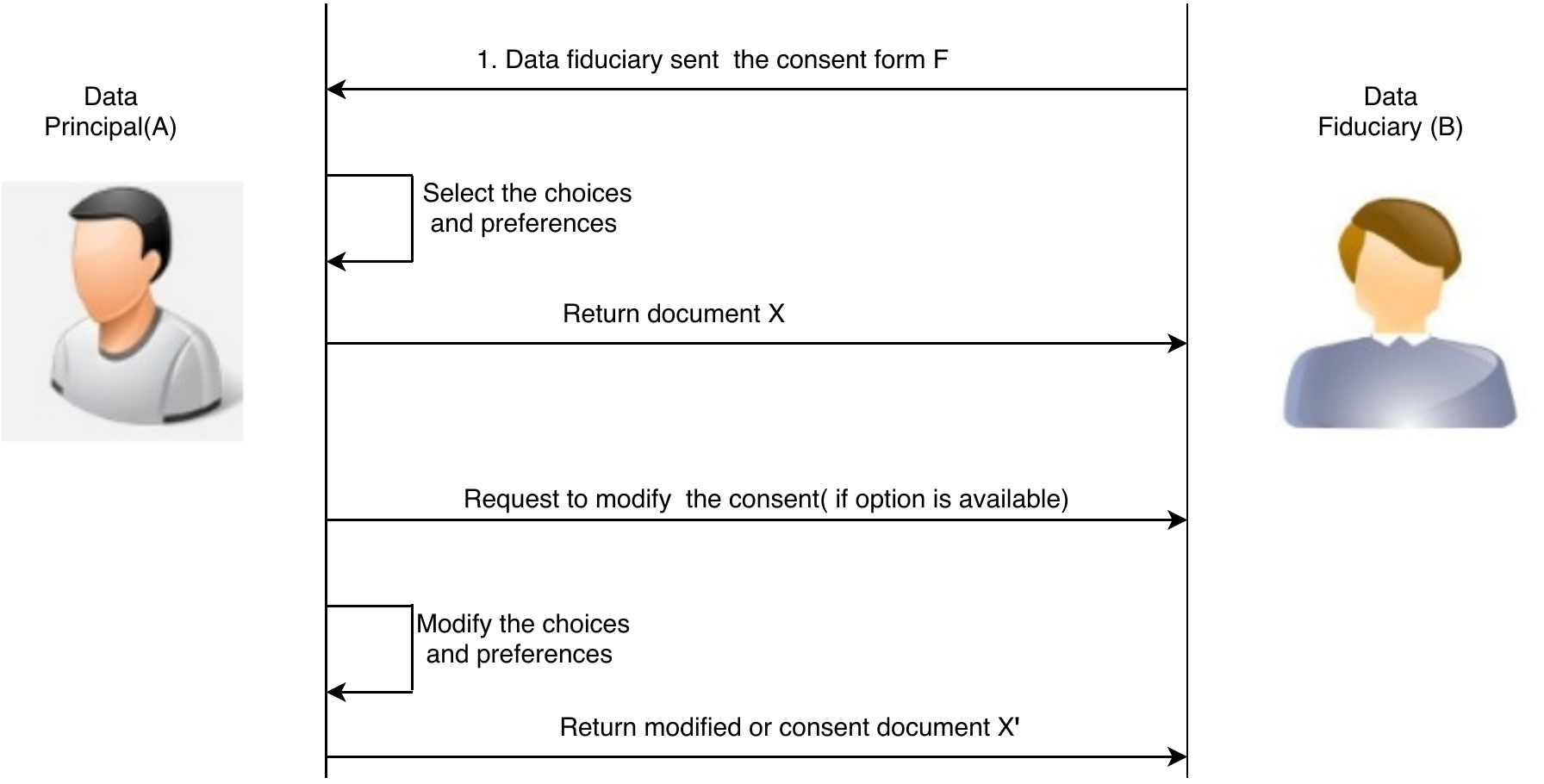}
     \caption{Current model of consent establishment  between data principal and data fiduciary }     \label{fig:poc}
\end{figure}

 \textbf{Existing methods of retrieval of consent and its limitations:} The current status is, organizations are trying to modify their business so that they can provide more choices, preferences and control of data to the users in the form of  consent. Figure \ref{fig:poc} shows the current model of establishment of consent. Data fiduciary  provides a form $F$ to the data principal. Generally, data principal gets either of the two options. In the first one, he has to accept or reject the consent. In such cases, all  terms and conditions are enclosed in the consent form itself.  In the second case, data fiduciary provides opt-in methods which allow users to select few choices and preferences. After selecting appropriate preferences data principal can return the document $X$. Few organizations allow to modify the consent. To do this data principal can request to modify his choices. He can re-submit it as document $X'$ after doing necessary changes. 
 
 \par
 
 The current model of taking consent does not fulfill the properties of consent discussed above. It has several limitation and drawbacks. For example, both  accept/reject method and  opt-in method of collection of preferences  does not guarantee correctness of proof of consent(PoC) establishment. There are three reasons: 
 (1)Since consent is  managed solely by data fiduciary itself it may allow data fiduciary to perform maliciously by altering the consent form. It can also prove to the individuals(?) that consent has been taken even without obtaining it.
 (2) Even if it takes consent, there is no way of verifying that processing was done according to the terms in the consent form. 
 (3) processing is done on the basis of recent (modified) consent or old consent. 
The current model also does not provide an option to  delete the consent. So, there is a need to  implement proofs of consent properties which can resolve dispute and breaches.  

To achieve the goal of consent, the data fiduciary needs to implement all the items listed in  Subsection \ref{subsec:proofofconsent}. We have provided a high level  overview how technical methods can used to implement this. We have skipped the detailed construction  as a future work. If we assume that data fiduciary uses opt-in method of consent then items 1, 3 and 8 of subsection \ref{subsection:proofofconsent} can be implemented directly by writing all the conditions explicitly. For this, data fiduciary can create a simple form with opt-in methods. All purposes, list of third parties and explicit consent  for data processing can be mentioned in the form thoroughly. 

%(DIFFERENCE BETWEEN 1 \&8)

\par 
\textbf{Undeniable consent proofs:}
 We need to address how can we encode the consent into the data collection and storage software to satisfy Item 2. One way is using \emph{digital signature \cite{DSA}} schemes which will work as non-repudiation technique for the consent. Data principal should digitally sign the consent form. Signature on the document will ensure two things. 1) Data fiduciary can prove that he has obtained consent before data processing and data principal can not deny it.  2). No party other than data fiduciary can have such proof of consent (if implemented correctly with combination of other cryptographic techniques).  To modify the consent (points 7) or withdrawal (point 6)  of the consent, data fiduciary can ask user to resubmit another form with different signatures.  Therefore, signature can help to resolve the dispute of consent establishment between both parties.   
  \par 
  %right to withdraw part of point 5 and point 6
 \textbf{Inclusion of time information:} In order to prove ``since when'' parts of point 6, digital signature is not sufficient. Signature can not guarantee correctness of time since when consent is established, modified or withdrawn.  To prove when the document was signed, any secure time stamping \cite{timestamp} algorithm can be used. 
 
 \par
\textbf{The dilemma of old and new consent proof:} Items (5, 7,9,10) describe the cases of modification or withdrawal of the consent. Modification in the consent can be started from either of the party. 
It is possible in the case if any change happens either  due to the  privacy policy of the organization or change in policy of  PDPB standards.
%When data principal wants to modify his preferences, he can  a request for the change. Similarly, data fiduciary can also do  request to the data principal to modify the consent. 
As discussed above, data principal can submit modified consent form along with the new signature. Now, the data fiduciary has both old and recent copy of the consent. Both consent are valid and processing can be done upon either of them. This may allow data fiduciary to be malicious. For instance, he can delete either copy of the consent and can do processing on the basis of his choice of consent. In such cases, data principal can not prove malicious behavior of data fiduciary. Therefore keeping both old and new consent does not enforce data fiduciary to process the personal data based on recent consent. 

\par 
To distinguish  between old and new consent  blockchain \cite{blockchain} based protocols can be used. Blockchain  is a  tamper proof, immutable verifiable ledger used to record transactions. Transactions recorded on the blockchain are transparent. Proof of consent can be recorded in any public blockchain for example,  \cite{ Ethereum}. Since all the consent will be available on public chain, both party can  agree on the latest consent. This will also ensure parties to resolve the dispute (if any) through the legal authority. We will provide a formal construction of consent properties in a follow up work. \\ 

%Add reference of consent and blockhain  from ICBC 2020
%Consentio: Managing Consent to Data Access using Permissioned Blockchains, Rishav Agarwal, Dhruv Kumar, Lukasz Golab, Srinivasan Keshav and Put Your Money Where Your Mouth Is - Towards Blockchain-based Consent Violation Detection, Jonathan Heiss, Jacob Eberhardt, Max 
\par
\noindent \textbf{Other desirable properties of  proof of consent}

There are few consent properties where proof of correctness would be difficult to achieve, but these are desirable. For instance, data fiduciary collecting data without consent (Item  11),  or data fiduciary is sharing data with unauthorized  parties (not mentioned in the consent form)  (Item 4),  or data fiduciary is processing data even after consent has been withdrawn (Item 5). We summarize more such desirable proofs  below: 

 \begin{enumerate}
\item
  Data fiduciary should implement transparency in disclosure of the list of third parties with whom data is being shared. On the other side, as per the bill, third party has to disclose the source of data, if it is not collected directly from data principal. For this, he should implement a proof to show the origin of personal data. Such proofs are necessary because no one actually knows how data is processed behind the wall \cite{7sin}. 
  
  %NEEDS DISCUSSION??
  
 \item  Third parties  should  implement proofs that  processing is being done as per the contract with data fiduciary. 
 
 \item If consent is being withdrawn by data principal, same will also  be communicated to third parties. Now third party should implement proofs that no processing is being done after consent has been withdrawn.  
 
 \item Data fiduciary should implement proofs that he is  not using data for other purposes. If he wants to do so, he 
 he can inform to the data principal and can ask separate proof of consent.
  As of now, such proofs are difficult to implement and would required more advanced methods. 
 
  \end{enumerate}
Currently, technical solutions to  complying with consent have not been  discussed thoroughly in the literature. 
Though many authors claim \cite{BC36_blockchainPDP, BC4_HealthCareDataGateway,  bc10_1_DataStorageIoT} that Blockchain is a panacea for this, the authors are skeptic. 
The primary reason is the difficulty to prove what operations are performed on data. 
Maintaining an audit log is not sufficient, as it cannot keep track of activities performed by unauthorized users. Even data fiduciary might choose not to log some of the events, in particular when it shares data with unauthorized third parties or performs illegal operations.  

%NEED TO VERIFY THIS: 
 %Therefore, in future each  should be analyze and solve independently??.
 \par
 The more advanced technical method could be explored to solve the existing challenges of proof of consent. For instance, to verify the consent a more advanced cryptography based  method such as \emph{third party audit \cite{TPA1}} could be developed.  The audit scheme will increase transparency and minimize manual interference. The algorithm may check and verify multiple parameters to confirm that consent is as per the PDPB standards. From legal perspective, a penalty can be imposed if it is found that either data fiduciary or third party is processing data for additional purposes without prior consent.

 \subsection{Data collection} \label{section: datacollection}
Data can be collected after retrieving  the consent from data principal. In the bill, many sections cover the properties of data collection such as Articles  5, 6 and 9 corresponds to  the \emph{``purpose limitation, collection limitation, and storage limitation''} respectively. From implementation perspective  the following properties should be considered to prove that data collection is as per the regulation.
 
  \begin{enumerate}
      \item Purpose limitation
      \item Collection limitation(by following minimum data collection, and putting restrictions on category of personal data being collected)
      \item Storage limitation (duration of personal data storage)
      \item Information about third parties with whom personal data is being shared.
      \item Security and safeguard of  data (while collection, transfer, share and  storage of data)
  \end{enumerate}
  
  Data fiduciary should identify and inform  the purpose for which data is being collected \emph{(purpose and collection limitation)}, and disclose estimated data retention period \emph{(storage limitation)}. If such period could not be determined then disclose a tentative duration. To compute this, a review can be conducted on regular interval, and can be informed to the data principal accordingly. Similarly, each data fiduciary has to maintain the list of \emph{``third parties with whom data is being shared"} (as per  Article 7).  This list could be made available at their website or can be informed directly. Further, during collection, cat fiduciary should disclose other information such as \emph{``any cross border transfer of data''}, \emph{``type, nature and category of personal data being collected''} or \emph{``source of collection if data is not obtained directly from data principal''}. All such notification should be done before the commencement of processing. 

%We discuss here how the goals of data collection can be achieved? 

 %We have discussed collection of sensitive personal data in separately in  section X.
 
 %How this \emph{minimum}  There is no unique model for it. It will  In particular,
  
 %\subsubsection{ \emph{The Primary goal of data fiduciary is to collect data least and minimum}} \label{subsection:primary goal}
\subsubsection{Goal of data fiduciary: collect minimum data} \label{subsection:primary goal} 
  
The primary accountability from data fiduciary is, he should prove that collected data is minimum. It is necessary because data fiduciary can have  strong monetary incentive by collecting enormous personal information. Sometimes sensitive data like health  information, credit card number or medical record  \cite{smartphone_UL_MDC,appselling_FB_MDC} are collected even without user's knowledge. But, it is still a big question, \emph{how it will be justify whether collected data  is necessary and minimum?}. We discuss two approach here that can provide a direction how technically data fiduciary can minimize the collection. The first approach is, analysis of what to collect using flow chart.

 %He should use this data only for specified purposes. 
%which data fiduciary  needs to consider while designing  minimum data collection and  to accomplish the goal of data collection.
 %Why \emph{data fiduciary needs }. 
%Data fiduciary should analyze the requirement and collect only the justifiable amount of personal data as well, data fiduciary will use 

%any point of the time data fiduciary finds ,
%\emph{data fiduciary must avoid unnecessary collection of personal data}.  Data fiduciary has to  ensure that  data is being collected for one purpose shall NOT  be used for other purposes. 

%The incentive  will be proportional to information that data fiduciary collects. 
 %Due to all this, PDPB imposes a restriction that data fiduciary shall collect only \emph{minimum, justifiable personal information}.  
%Proof of minimum collection is depends on the type, nature and scope, kind of services offered by the organization and the way  data is processed. He should gather only the  necessary and justifiable data. 
% He has to gather only the  necessary and justifiable data and,

 \par 
  \begin{figure}[!ht]
     \includegraphics[width=\textwidth]{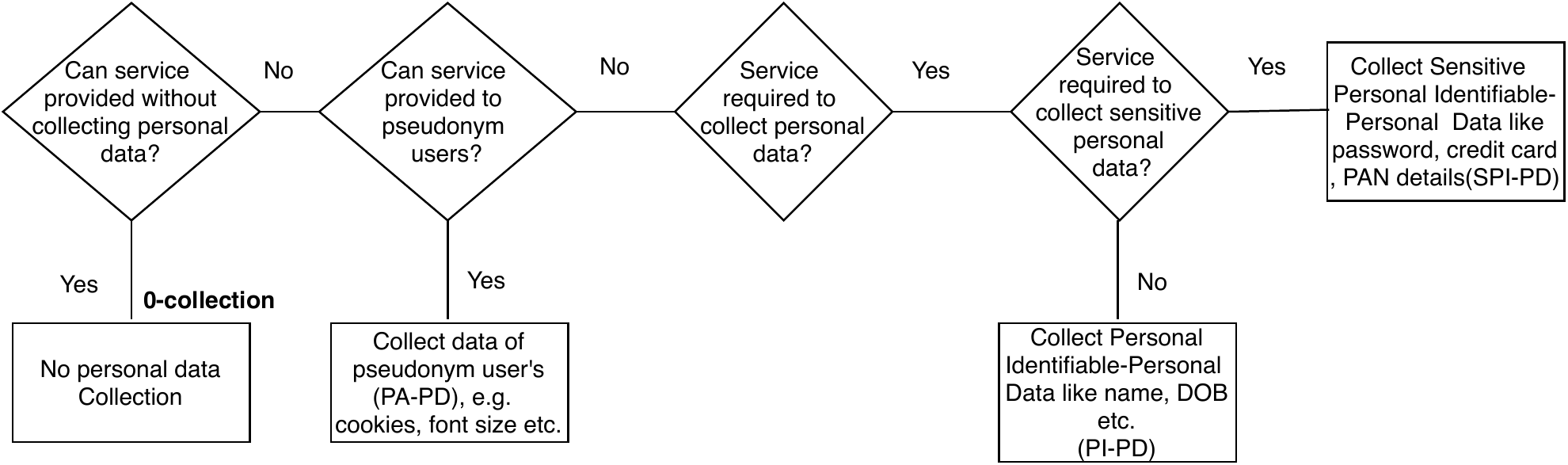}
     \caption{Data collection minimisation using flow chart} 
     \label{fig:dcfd}
\end{figure}
  
 % We describe here how an organization can take a decision about what data it should collect.  How it will be minimum and achieve the design goal of data collection mentioned in the point 1-5. 
 
 \textbf{Minimisation using flow chart:} If data fiduciary can decide what  NOT to collect, then he can achieve a significant level of the goal of data collection. One model that can help here is flow-chart. Using this, data fiduciary can decide whether or not any information he should ask from users. For example, at very high level data fiduciary can create a flow chart of four kind of choices (as demonstrated in Figure \ref{fig:dcfd}). First, if data fiduciary finds that, ``service can be provided without collecting any personal information then he must not collect anything''. We call it \emph{zero-collection}. In such case, organization must disclose their privacy policy of zero collection. He can do this either by disclosing the policy directly to data principal or by prompting the message at their website. At practical level zero collection can be implemented by organizations who provides informational services, user awareness services or   knowledge sharing services \cite{wiki, Noprivacy}.

 %If not, he must not collect it. Otherwise he should collect SPI-PD
 
 \par 
 Data fiduciary should obtain further information  such as IP, cookies, session  or browser information. These information are set for pseudonymous users (known through nick name over the internet). Any physical identifiable information should not collected at this stage. We call it \emph{pseudonym associated personal data (PA-PD)}.   Consent and choice of preferences will still required  before such collection. 
 At next level, data fiduciary can go for the collection of \emph{personal identifiable  personal data(PI-PD)} and \emph{sensitive personal identifiable personal data(SPI-PD)}. PI-PD are like name, date of birth or mobile number. SPI-PD are like financial information, health data or  biometric data. Data fiduciary should analyze thoroughly which one should collect from both.
 
 %prove that whether their collection of sensitive personal data is necessary?  

  \textbf{Enrichment in data processing methodology:}  The second approach of minimum collection is, data fiduciary should change their existing data processing methodology. A few technical changes in the processing will not only full-fill the goal of  data collection but also serve the same business purpose. We have justified our argument using two use cases. Both use cases conclude different observation. First use case argues, the requirement of technology change in the processing, while second use case argues, the necessity of collaboration among data processing parties. Since data fiduciary is collecting multiple sensitive personal information in both use cases. We have argued that neither  it is minimum data collection nor it is providing data protection. Later, we have shown that the same objective (the goal of minimum data collection, and necessary data protection)  can be achieved by just a few modification in their processing activities.
  
  \subsubsection{Use Case 1: storage of debit card/credit card information}
\begin{figure}[!ht]
     \includegraphics[width=\textwidth]{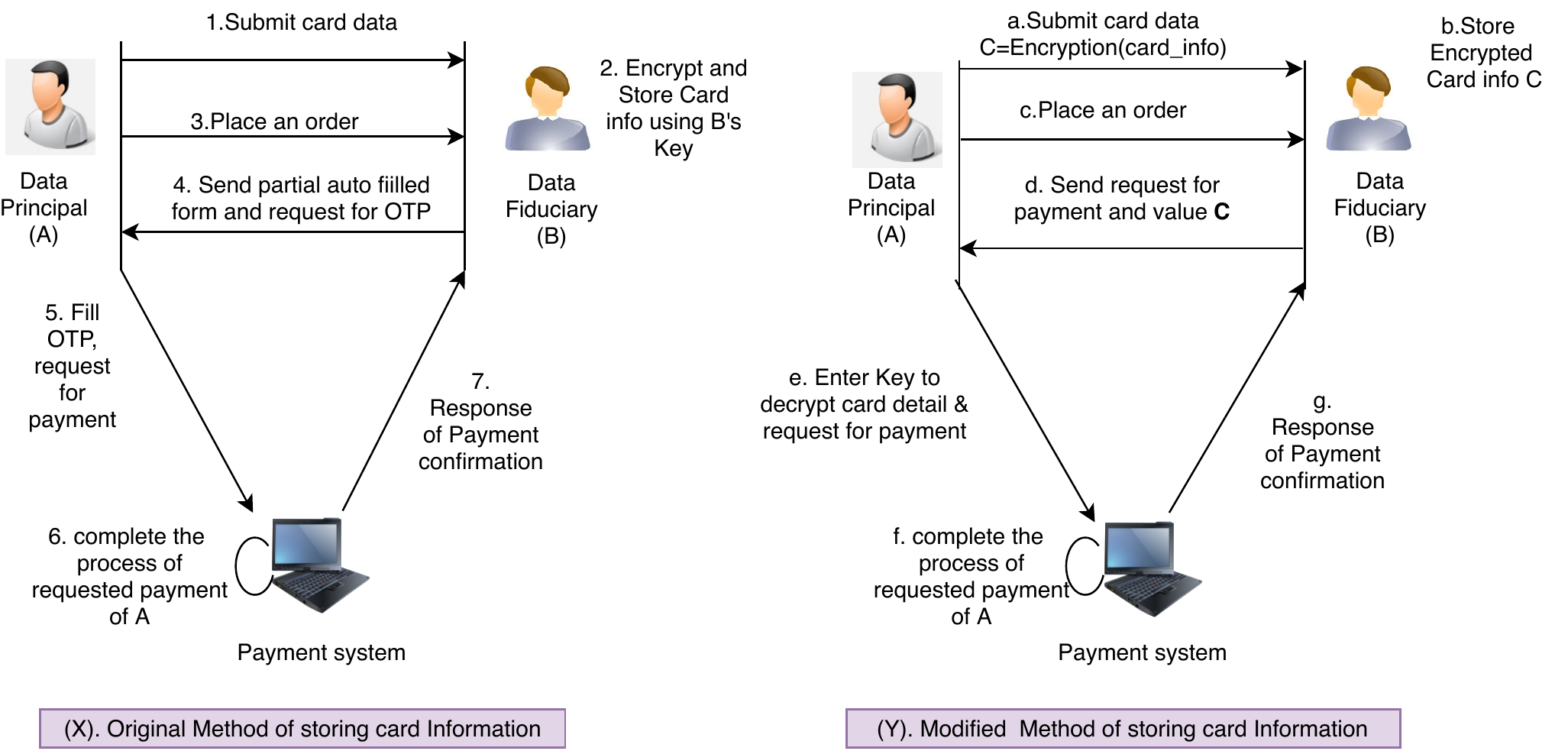}
     \caption{An example of data collection: original vs modified method }     \label{fig:spi}
\end{figure}

  Debit or credit  card data is  sensitive personal information and requires extra protection during processing. Commercial platform  provides \emph{ease of doing} facility wherein quick payment could be done by storing user's card details \cite{flipkart} at data fiduciary portal. The  purpose is stated as: \emph{``It's quicker. You can save the hassle of typing the complete card information every time you shop at Flipkart by saving your card details. Your card information is 100 percent safe with us. We use world class encryption technology while saving your card information on our highly secure systems''} \cite{flipkart}. 
  Fig. \ref{fig:spi}(X) shows the current model in which data fiduciary stores card's information to facilitate quick transactions. The data is generally encrypted by data fiduciary's key hence always available to him. Whenever user's place an order, merchant sends a payment request. Payment request contains partial auto filled form having card information already filled. Thus, user's does not need to submit card details manually. In the next step, user enters one time password(OTP) and submit it to the to the payment system. The advantage of such processing is user would get ease of doing and quick processing by not entering card information manually.
   
  \par
  After introduction of bill, it is questionable whether such collection and storage is required? If yes, then in which form data fiduciary should store. Because, data fiduciary is non trustworthy,  and  weak protection of data may reveal card detailed publicly or  card information can be used for malicious purposes. Data breaches over here may lead to severe harm which sometimes appears in the news when user's card information becomes available over the the dark net. Hence such storage increases doubts on security and privacy of user's data. Whether the same purpose \emph{quick transaction and ease of doing} could be achieved by preserving privacy, protecting card details, ensuring limited data collection and the data minimization?
  To achieve this, each party involved in the processing should get necessary information only. Data fiduciary does not need to know the details of card. It should be visible to the bank only. As the specification of secure electronic transaction(SET) also \cite{SET} states, ``order information(OI)  information shall be processed by merchant and payment information(PI) by the bank''. Both section should be processed separately.  
  
  \par 
\textbf{Use of more advanced methods:} Cryptographic technique can help to achieve the above goal. For instance, figure \ref{fig:spi}(Y) ensures the privacy preserving minimum data collection, and also provides \emph{ease of doing}. 
  Data fiduciary collects card details as a cipher text $C=Encryption(card\_info)$ encrypted using customer's key. The key may be anything such as user's password. Since data fiduciary does not know key user's key so it is useless at the merchant's end and would be  available at the bank's end instantly. Whenever user places an order  data fiduciary can send the request of payment along with order details and  the cipher value $C$. User can decrypt card information and can forward  payment request to the payment system. Since decryption is being done at  user's side  data fiduciary would  not able to know the card  details. It will ensure the goal of data minimization, purpose limitation, collection limitation, storage limitation and sharing of information.

\par
 
 \begin{enumerate}
     \item \textbf{Purpose limitation}: In the first case, merchant stores debit/credit card information directly which is not safe. While in the second, storing encrypted card information, does not reveal any information, serve the propose of ease of doing and such details can not be used for other purposes. Hence, it  achieves the purpose limitation.
     
     \item \textbf{Collection limitation(data minimization)} Storage of multiple card details can reveal multiple information. In the second case,  card information is not available to the data fiduciary which limits the amount of collected of data.
     
     \item \textbf{Storage limitation} By using encrypted storage, user does not need  to worry about data retention period of sensitive personal information.
     
     \item \textbf{Sharing of data} No meaningful information will revealed by sharing  card details to the third party.
     
     \item \textbf{Security and safeguard} Encrypted storage ensures security and safeguard of the data at fiduciary's end. 
 \end{enumerate}
 
 Data protection bill encourage that data fiduciary should implement such kind of  technology. It would enforce visibility of data at the right place only and, with appropriate security and safeguard. Uses of more cryptographic methods  will  enhance the confidence of data principal that he has control upon their data  and it is safe. Such techniques would also provide transparency during processing, storage and sharing of data.
 
 %This kind of data collection goal  may also provide  data fiduciary to build a  trust and confidence among user's that  data which has been collected, transferred and stored are secure and it is not affecting the privacy of data principal. 
 %In future we need to develop more CRYPTO methodology which shall be require to prove  various aspect of data  protection framework and achieving the goal of data privacy. 
 %Limited data sharing and prevention of disclosure of sensitive  personal information 
 
 \subsubsection{Use case 2: Prevention of disclosure of sensitive  personal information  - a PAN card example }
What if one data fiduciary is enhancing their processing methodology while others are not?  Sometimes it requires the participation and collaboration of all parties  involved in processing to upgrade their techniques. Through the use case of PAN card we have shown that until and unless all the parties with whom data is being shared does not enhance their data processing methods the goal of data protection can not accomplished.  
 
 %the   This use case  proves that \emph{achieving the goal of data protection by modifying the processing methodology is not the solo responsibility of individual organization}. Sometimes it is need  They also needs to modify their business collectively.
 
 \par
 
\begin{figure}[!ht]
     \centering
     \includegraphics[height=8cm, width=0.8\textwidth]{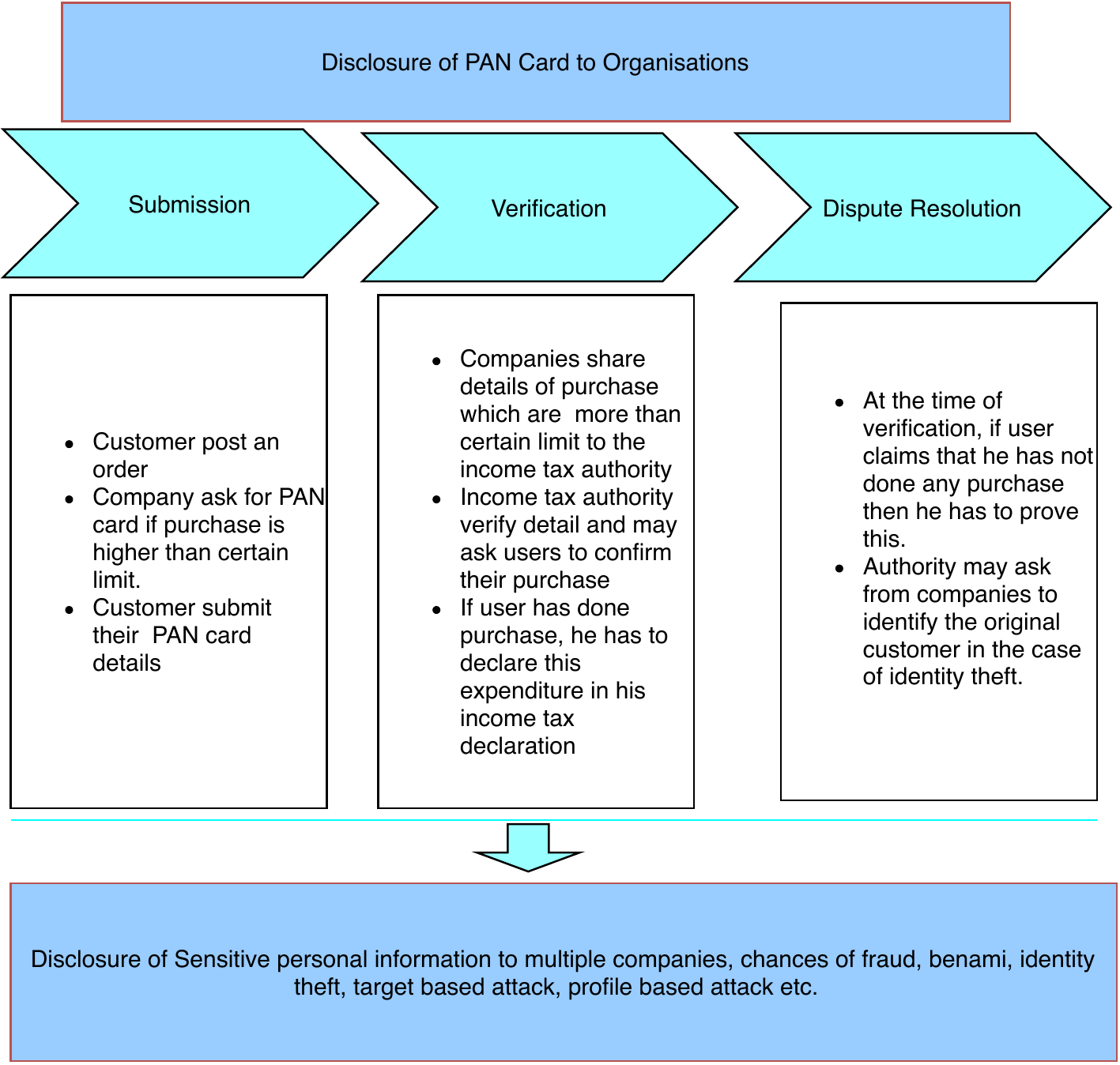}
     \caption{Existing model of sharing of PAN information}     \label{fig:panold}
\end{figure}

  PAN card is a unique number assigned to all tax payer within India. Income tax department keeps track of individual's tax declaration and their income. Is can also be used for identity proof. If any one does purchasing or availing a service of  of  more than specified amount he  has to  disclose the details of PAN card to the merchant. The authority can use these details to inspect individual's tax declaration. It has been observed that details of PAN card collected by merchants can be used for malicious purposes  such as to purchase benami properties, for identity theft \cite{PANtravel}, to perform fraud payment \cite{PANfraud1} or to hide income taxes.  Consider one example of such PAN detail submission at the  commercial websites  as shown in figure \ref{fig:panold}.  Merchant collects PAN details when user's does purchase of more than specified limit.  Merchant forward information of purchased details to the IT department.  The IT authority may verify  whether or not purchase details violets the regulation of tax disclosure. In the case of fraud/theft, an investigation may be started against suspicious user's.  
  
  % For instance, individual presents PAN details as identity proof at various places. Disclosed PAN details can be used by someone else to declare their expenditure of higher amount of the money  using identity theft. In this case, it would be difficult for the real user's to prove that he has not expend higher amount.
  
  %We discuss here the present model  of  PAN card details by the merchant. We have shown that  how this ineffective collection of  sensitive personal information can be used for malicious intention. Such collection also does not complete the goal of income tax authority to monitor tax payer's purchase activity. Later, we have shown that how a minor change in the present  model may lead to achieve the goal of PAN detail  collection, fulfill the requirement of income tax , as well it can achieve other properties of data collection.  

  %of product is  indeed done by customer or not. If yes, then customer has to include the purchase details in his income tax declaration. 2). If it is the case of identity theft, then authority may ask customer to prove that he has not done purchase. If customer is not able to prove, a fine can be imposed on the customer. 3). In further analysis, tax department can try to find the customer who performed identity theft and done the purchase. 
  
  %For this, authority may ask  merchant to identify original customer who has used some one else PAN card for the transaction, so that punishment  can be imposed. 
 
 The existing model does not solve  the intended  goal (transparency in expenditure and declaration of taxes) of data collection. In this case multiple breaches are possible such as tracking of user's behavior by merchant, or the  use of card details for  impersonation and identity theft. From authority point of view details shared by merchant are not trustful. Merchant can share inaccurate data or he can share false information  submitted by the users. The authority expects more monitoring and transparency in the purchase activity. But, current model of PAN details submission neither solve the purpose of any party nor full fill the goals of collection.
 Further, even one party such as merchant enhance their methodology to full fill the goal of data collection still the purpose will not be solved until and unless other parties does not collaborate.   
 \par
 
 %Here the purpose of collecting PAN card detail is: \emph{to monitor the purchase detail of tax payer by income tax authority  so that transparency can be maintained in the tax declaration of tax payer}. Hence collection of  PAN details neither solve the purpose of data collection, nor the goal of income tax authority. This method of processing of PAN data does not preserve properties like limited data collection, limited data storage and limited data sharing  because merchants do not need to collect, store and share the PAN card information. Also, current method of sharing purchase details of customers  does not authenticate users (any one can create an account and use some one else PAN card number), does not provide transparency and privacy. Transparency means income tax authority does not have clear methods to monitor purchase record. Table  mentions such properties.
 
 \begin{figure}[!ht]
     \centering
     \includegraphics[height=8cm, width=0.8\textwidth]{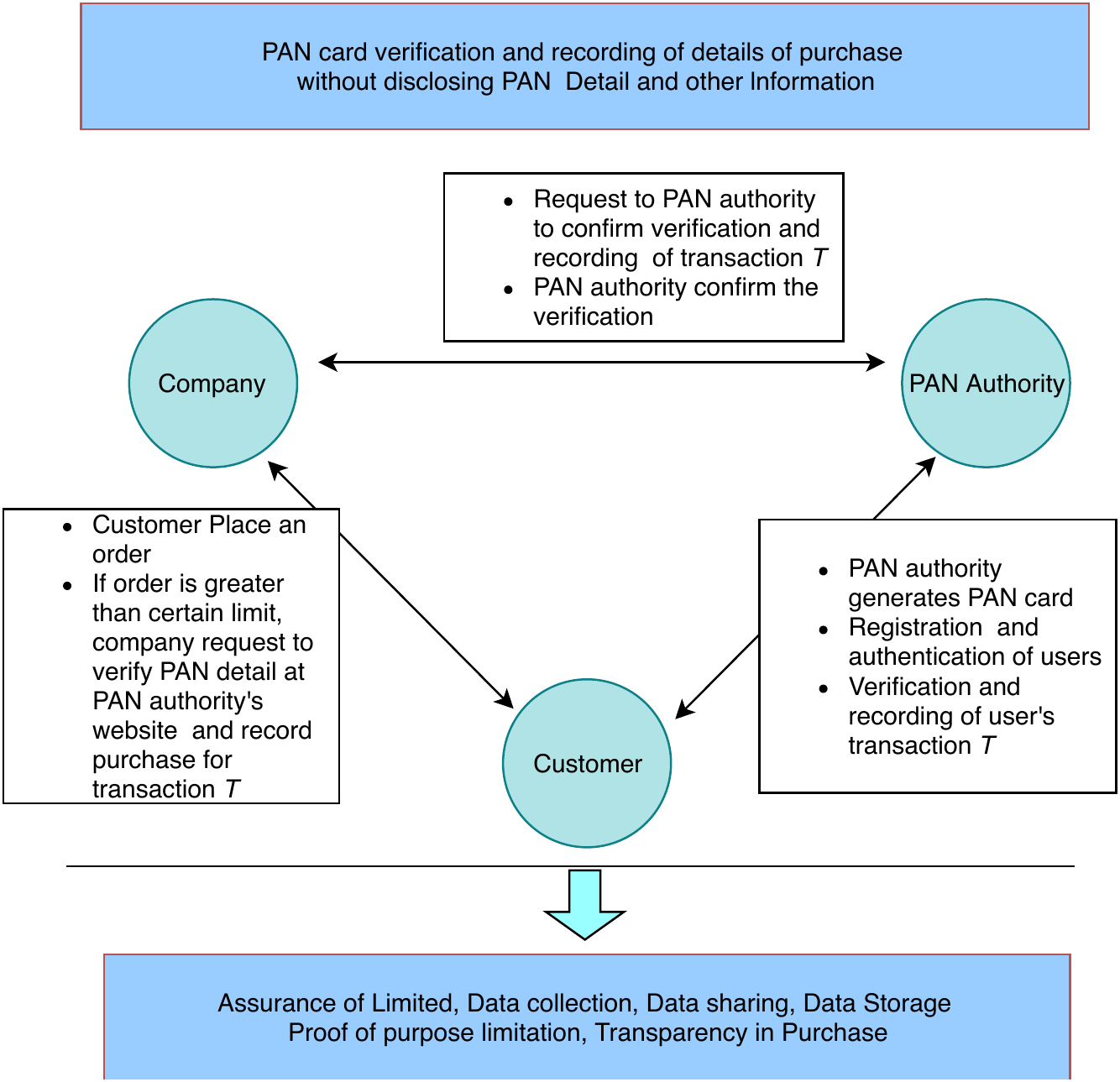}
     \caption{Modified model of sharing PAN information}  \label{fig:pannew}
\end{figure}

  \textbf{Collective change in processing methodology} Figure \ref{fig:pannew} shows that if all parties collaborate and modify their data processing approach they will solve the above issues. In this case, instead of collecting PAN detail directly data fiduciary(merchant) ask user's to logged their transaction (purchasing activities) $T$ at the IT portal if it is higher than specified amount. As shown in the figure both User
  and merchant can confirm logging of purchase details using appropriate authentication and verification. A comparison is shown in the Table \ref{table:pan}.
  In the new model every party is getting only relevant details. 
  Merchant is not collecting any PAN details hence achieves the goal of limited data collection, storage. Merchant does need to share anything (limited data sharing). All the required purchase are being logged by the authority using suitable authentication hence IT department can achieve transparency and monitoring in the purchase. Moreover such transaction will also eliminate the possibility of impersonation and identity theft and tax fraud.
  \par 
  Thus, using cryptology, enhancement in processing techniques,  modification in  existing data processing methods, and participation of every organization can accomplish the goal of data collection.

  %himself  at the IT portal and subitcustomer receives verification request along with a transaction id . He can authenticate himself to the IT dept portal and  can forward purchase verification request. At the same time merchant send  a request  to IT department validate the customer with transaction id $T$. The department record the transaction, purchase amount and confirm back to both parties.
  
  %party invlove in the transaction has to modify their it will be the combine responsibility of every party involve in the processing to modify their policy collectively in order to ensure that their processing activity is  as per the data protection framework. 

\begin{table}[htbp]
		\begin{center}
		   	
			\begin{tabular}{|l|l|l|}
				\hline
				\textbf{Property} &  \textbf{Present method of PAN } & \textbf{Modified method of PAN} \\
				 & \textbf{ details processing} & \textbf{details processing} \\
				\hline
			      Limited data collection & \xmark & \checkmark \\
			     \hline
			     Limited data storage & \xmark & \checkmark \\
			      \hline
			     Limited data sharing & \xmark & \checkmark \\
			    \hline
			     Authentication of user's & \xmark & \checkmark \\
			     \hline
			     Transparency in processing & \xmark & \checkmark \\
			     \hline
			     Privacy of personal data & \xmark & \checkmark \\
			     \hline
			\end{tabular}\caption{Comparison of existing vs modified model of PAN details processing} \label{table:pan}  
		
		\end{center}
	\end{table}
  %Achieving the goal of data protection, like data collection, data sharing etc  is \emph{collective responsibility}. 
  
  %\subsubsection{Other goals of data collection}
  %Collection, transferring and storing will require security and safeguard. While collection and transfer, data fiduciary should use a secure HTTPs channel. Data fiduciary should ensure confidentiality, integrity and authenticity of data during overall processing. To achieve this data fiduciary  can use standard encryption scheme, signature scheme, message authentication scheme etc. Safeguard measures must be  followed for protecting data while \emph{data is in move} or \emph{data is  in rest}. In particular, safeguard must be followed from the starting of the collection till the personal data is available with data data fiduciary and deletion of that data.
  %In the case, if  data has been  shared with third parties, data fiduciary will maintain  a contract to ensure the security and safeguard of data need to be  followed by third parties. Personal data shared with third parties shall also be bind by the consent user has given to data fiduciary. 
  
  %Some other methods also exists which may help in reduction in data collection and security and safeguard of the data. For example \emph{data minimization techniques}. 

  %We can say that several crypto methods can be implemented to ensure that data fiduciary can provide the service with minimum personal information.

 \subsubsection{Technological limitations of data collection}
  % isme auditor part and authority part ko heading bana kar likh do
  
  But the challenging task is how will be decided that collected  data by data fiduciary is minimum, justifiable, and  having limited purpose. Regarding this a technical proof would be required for data collection.  We summarize all such properties as follows:  
   
  \begin{enumerate}
      \item How one will prove that data fiduciary  is collecting limited data.
      \item How one will prove that data fiduciary is not using data for other purpose.
      \item How one can prove that data is used only for specified purpose(limited purpose).
      \item If data is processed for additional purpose, how will be proved that consent is received from users.
      \item If data is collected indirectly,how will be  proved that he has informed to data principal.
      \item Proof that he has conveyed to data principal regarding intimation of  any cross border transfer of data.
      \item How can be proved that data is being used for specified period. 
  \end{enumerate}
As of now the exact solution of above problem is unknown. One solution could be, data fiduciary can make data collection process more transparent. For instance, he can publish necessary reasoning and explanation for such data collection on their website and can inform to data principal as well. these publishing can be audited by DPAI.  Data fiduciary can also publish a \emph{certificate} issued by  DPAI. Certificate will work as a proof that data collection by data fiduciary is appropriate.  On the other side, if any one finds maliciousness in data collection means, data collection is not as per the standard then he can do complain to data protection authority(DPAI). Authority may setup an independent analysis or audit for against such incident. The Auditor can analyze over all collection activity and  the intention behind such collection.  The authority can impose punishment if data fiduciary found culprit. But these methods are just like safeguard, not a full proof methods. A further research and more advanced methods are required to solve the above problems.

  \subsubsection{Data retention policy: storage limitation}\label{subsection:storagelimitation}
    Data fiduciary would require to store the data for the period as long as it is necessary (as per Article 9). If such period can not be determined at the time of collection then a periodic review could be conducted to estimate the retention period and a tentative duration can be  provided to the data principal. The prime concern is, in the real world it is quite debatable and difficult to determine the period for which data will be stored. Presently organizations like Facebook, Google's are storing data for longer period than it is necessary for analysis and monetizing purpose. The essential concern is what should be the criteria for the organization to determine a duration to retain the old data. For instance, Facebook has stored the data since the user is active \cite{FB} while Whatsapp \cite{whatsapp}  deletes the data from their server once it gets delivered.  
    
  %\par 
  %Because it has seen in recent incidents.that  data fiduciary are not interested for the erasure of data. Consider the case of Indian mobile service provider who has collected user's Aadhaar details to provide SIM. Later, the apex court has stated that any private data fiduciary are not eligible to retain and process Aadhar details of an user and have been asked to erase such data . As of now, it is questionable whether  companies still holds and processing  such data  or Aadhar details have been erased from the system.
  \par
  \textbf{Inherent constraints: ease data hiding} how it would be proved that data is not being used after retention period over or data fiduciary has not kept another copy of data after the deletion \cite{SC}.  
  \emph{Proof of storage \cite{ProofOfStorage}} is one such cryptographic technique which can used to proof   whether data fiduciary has retained personal data.. With proof of storage integrity of remote file can be verified. Similarly, data fiduciary can implement erasure based proofs. Through it, one can verify that  data is erased and re-written in the volatile memory. Data principal can  request for \emph{proof of secure erasure(PoSE) \cite{ProofOfSecure_Erasure}} from data fiduciary to verify and validate data has been erased (re-written). However, these methods are not well formed and have their own inherited technical limitations. For instance, if data is copied at make hidden at another location then such proofs will not work.  Even, it  can not guarantee that data is not being used for further processing once retention period is over. From implementation perspective, a more practical methods need to develop to accomplish the goal of storage limitation.

 \subsection{Data Processing} \label{section:dataprocessing}
  Data processing consist of operations such as addition/deletion/alternation, analysis, sharing and storage performed over the data, to obtain, transform or classify the information. 
  %After data  is collected by data fiduciary, it  always associates with any of the stage of data processing. 
  As per the framework, data processing shall be done after the consent and data must be protected throughout all the significant steps of the processing. Although, there are many sections in the bill regarding data processing, we have discussed only the three major sections from point of view of software implementation.  These are,  \emph{privacy and security safeguard}(as per Article 24), \emph{privacy by design}(as per Article 22) and \textit{transparency}(as per Article 23). Security and safeguard is discussed in next Section and the other two, privacy by design and transparency have been discussed in the next subsections.

 %Such processing shall be  \emph{fair, reasonable} (as per section 4) and \emph{lawful}(as per section 7).%data fiduciary shall implement appropriate security safeguard. Effective incorporation of security safeguard, privacy by design and transparency  will lead to achieve the goal of data processing.

 %Many works have been done to achieve the goal of above. here
 %The big concern is  how technically it would be achieved because privacy and security  highly depends on the nature, scope and purpose of processing, likelihood and severity of harm of the data (as per Article 24). Due to this system to system implementation  of privacy, security and safeguard and transparency  could be different.  In the below section, we have provided  We have also discussed the clauses of the framework  where such techniques are applicable.
%We also discuss here why privacy by design and data transparency is necessary for data processing.  
%\subsubsection{Security and safeguard} \label{subsection: security and safeguard}
%Cryptographic and other techniques which empower protected personal data processing

 \subsubsection{Cryptographic and other techniques for protected data processing} \label{subsection: security and safeguard}
We describe the list of  security and safeguard goals mentioned in the framework.  For each goal, we have discussed the set of available  techniques  to implement it.  There are numerous methods present in the literature  that can provide a strong security and privacy of  data. We have discussed below, some of the cryptographic techniques. 

%Security is essential requirement for data protection. 
%Safeguard in the data protection can  be \emph{encryption, anonymization, de-identification, integrity or access control} of  the data. In following section we have described each one and asserted that  efficient implementation of each can provide guard against data breaches. 
\par 
 \textbf{Incorporation of encryption methods:}
Article 24.1(a) of the framework says, \emph{each data fiduciary needs to implement appropriate methods of encryption}. \emph{Encryption} is a very old method which converts plain text into cipher text. It should be used as a primary tool  at every significant step of data processing. There are two kinds of encryption: symmetric and asymmetric. \emph{AES, 3DES} are examples of standard symmetric encryption schemes \cite{mc} and \emph{RSA,  Diffie-Hellman key exchange}  are the example of asymmetric encryption schemes \cite{pbe}. 
 Encryption algorithms are theoretically proved to be secure and correct implementation  can provide confidentiality, security and privacy of data. 
 %yaha karna hai
 %But the above 
 %Based on nature of the system suitable encryption scheme can be implemented.
 
 %Such schemes has to be implemented by data fiduciary at both when data is moving or data at the rest.
 \par 
 Improper implementation of encryption mechanism may lead to data breaches. There are instances where breaches have been reported even when the data fiduciary claimed that they follow best encryption schemes and well defined practices for security of data.  For example, Aadhaar, the biometric based unique identification system for citizens of India was subjected to data breach \cite{breach_aadhar} due to poor implementation. Unauthorized access of data also resulted in data breach in \cite{breach_starwood,breach_heritage,jio}. 
 %Further, security and privacy definition of encryption methods does not withhold after  decryption of cipher text, or when data is improperly disclosed in plain text form.

 \par
  To achieve more security and privacy, data fiduciary can use multiple kinds of encryption at different levels such as \emph{user data encryption, file-system encryption, database encryption or disk encryption}. All these methods provide security when data is at rest. Based on the architecture of the system data fiduciary can decide which encryption technique would be appropriate. 
 End to end encryption mechanisms such as \emph{SSL/TLS}  should be used for all data communication.

  \par
  Correctness of implementation  of encryption schemes  can be verified independently by any third party audit team. Audit team can analyze which kind of  encryption methods data fiduciary is using and how well it has been implemented. They can also provide necessary guidelines to the data fiduciary if they find that the used procedures are not sufficient to protect the data. The encryption algorithms, with approprirate parameters should be made public to improve the transparency. 
 
\par
  \textbf{Client side encryption of cloud  data:}  The growth of cloud will be higher with the recent  development of advanced technologies like ``software as a service(SAAS) \cite{SaaS} and  software defined network(SDN)" \cite{SDN}. In future many data fiduciary will  shift their adequate number of computation, services and storage at  the cloud. For such mass storage, encryption of cloud data would be essential.
 But the main problem is  most of the cloud service providers uses \emph{server-side encryption}. In this, the key is owned and stored by service provider itself. Hence user's data is completely accessible to the cloud servers. Breaches in such data may be harmful \cite{Digilocker}, even data can be shared and used for malicious intention \cite{breach_cloud, Digilocker}.  
 \par  
 %\textbf{The breaches could be more devastating if cloud has processing sensitive personal data.} 
 %The storage of sensitive personal data at the cloud  can be more harmful. For instance, consider the case of \emph{Digilocker}, a cloud based wallet service  provided by the government of India  to its citizen \cite{Digilocker}. Anyone can upload their sensitive personal documents such as driving licence, passport or PAN card etc. These stored  data can be accessed and used maliciously by an attacker for  impersonation, identity theft or other illegal activities. Even data fiduciary (government) can analyze user's stored personal data. In such cases, user's would like to store and process the data on the cloud securely with an assurance that cloud owner could not access and analyzed its data. 
 
 %To protect data from cloud server and to provide stronger  privacy of user's data, 
 %\par
 
 %Security and safeguard of user's personal data should be clinched on clouds as well.
 \par
 To address the above limitations, data fiduciary should  encourage data principal to store encrypted data on the cloud. This can be achieved by using \emph{client-side encryption (CSE)} schemes. These are cryptographic techniques that encrypts data at user's end and transmit cipher text to the server. At commercial level such techniques are also known as \emph{``bring your own encryption(BYOE) or bring your own key(BYOK)''}. BoxCyrypter and Cryptometer \cite{Boxcrypter,Cryptomator} are example of two such application that implements client side encryption (BYOK) schemes. Data is believed to be secure because  master key is managed by user's itself. 
 
 %(SHOULDN'T THE DATA FIDUCIARY ENCRYPT THE DATA?)
 
  \par
  %MERGE THIS AND NEXT DISCUSSION ON INTEGRITY PROOFS
  \textbf{Encourage the use of integrity proofs:} With the enforcement of the law, all the data, messages and document used for processing would have legal accountability. Data fiduciary has to disclose a proof of: i) nature and category of data  being collected from data principal (as per Article 7(b)) ii) source of personal data if it is not collected directly from data principal (as per Article 7(f)), iii) the list of third parties with whom personal data is shared and disclosed(as per Article 7(g)) and iv)data fiduciary has to take the necessary step to protect the integrity of personal data (as per section 24.1(b)). Many of the above points could be implemented by providing an integrity proof of origin of the data.  This may include the proof of source of data, identity of the sender and properties of non-repudiation. Data fiduciary can prove legal accountability for the personal data using such proofs. 
  
  %proofs of integrity of data.   i) the identity of the sender who has send the message or document ii) message has sent by sender only iii) message can not altered once it is received  iv) proof should become invalid if the message gets altered.
  \par

\par 
 %\textbf{Implementation of integrity proofs}
 
  Integrity protection is an assurance of consistency and accuracy of  data during entire life cycle of data processing. It ensures that data fiduciary is honest with respect to their process and any malicious changes in the data during storage, processing, sharing and  disclosure will lead to loss its integrity. Therefore, a verification and validation of data is essential for integrity check. To do this, four kind of integrity validation could be done: i) message integrity (validation during transfer of data), ii) system integrity  (validation of  data storage), iii) assurance of no information leakage (during   sharing and disclosure of data), and iv) data erasure integrity  (proof of data deletion). 
 
 \par
  Integrity of message should be ensured during transfer of personal data. Regarding this, both the sender and receiver can implement \emph{cryptographic message integrity validation schemes} with \emph{authenticated encryption(AE) \cite{AE}} or standard \emph{message authentication code(MAC)\cite{MAC}} algorithms. These algorithms would ensure that the received data is in original form without any modification.   To ensure  integrity of data in physical storage, data fiduciary could use techniques such as \emph{ integrity validation of data storage, redundant data storage and  data storage at multiple geo-locations}. Similarly, data fiduciary could also perform \emph{file system level or block level check-sum} integrity verification. In such  methods, an initial check-up is performed against the file system or blocks to validate the integrity. 
 \par
 Meaning of data deletion  integrity is the honesty in the action of deletion. To prove this, data fiduciary can provide \emph{erasure based proof \cite{ProofOfSecure_Erasure}}  which will verify the  integrity of data deletion. Data fiduciary also needs to ensure the integrity of  personal data when data is shared, disclosed with  third parties.   However, currently no standard methods exist to prove that data fiduciary maintains integrity while sharing and disclosure because data fiduciary can make redundant copy secretly and use it for malicious intentions.  
 
 \par
 \textbf{Authentication of users and services:}
 It describes how users identify himself to the system. Improper implementation of authentication method may lead to data and security breaches. Attackers can modify, disclose or destruct personal data or can perform other malicious functions. As per the bill, \emph{data fiduciary needs to take necessary actions to implement authentication, ``to prevent misuse, unauthorized access to, modification, disclosure and destruction of personal data"} (as per Article 24.1(c)).  To implement the above, strong user authentication methods like \emph{two factor authentication \cite{TFA}, multi-factor authentication \cite{MFA}, continuous data authentication scheme \cite{continuous_UA}  and time based authentication \cite{Timebased_A}} can be used. Each kind of authentication method provides various degrees of certainty that right object is interacting with the system. 
 
 %The foremost step is to establish well defined user authentication. 

  \par
 
\textbf{Authorization and access control: No one can access all:} Apart from authentication, data fiduciary should implement correct  \emph{ authorization}. Authorisation defines who can access what resources. Not everyone should have access of everything. Unauthorized access may also leads to modification, disclosure and destruction of data. Authorization can be implemented by creating access policy. Based on nature, scope and context of  data, type of users,  nature of processing data fiduciary can create  \emph{conditional access or access control} for various objects. Examples of possible conditions are  OS, software version, browser, IP, geo-location etc.  
\emph{Access control}
  is important because many attacks are launched by unauthorized users \cite{breach_starwood,breach_heritage,breach_aadhar}. A well  configured access control can prevent  systems from data breaches and  ensures security and safety of system. Based on design of the system, a set of specific access control methods can be used like, role based access control \cite{RBAC}, access control list, policy based access control \cite{Policy_PBAC}, capability based access control \cite{Capability_CBAC},\emph{ attribute based access control \cite{ABAC}, geo-location based access control \cite{Geobased_GBAC}, geo-fencing and lattice based access control \cite{lattice_LBAC}}.  Data fiduciary must be able to that  identify and authenticate users accessing the data,  are legitimate and have access of personal data within constrained domain. 
   
  \par

  %Instead of using single kind of encryption to store and protect the data, implement multiple encryption method at different level when data is at the rest. 

 %To protect the data over cloud, to provide Encrypted data storage at cloud \cite{encryptedcloudstorage}. 
 %Because usually major cloud storage service provider stores their data securely but they have complete access of it. They can perform search over it, do analysis, and even data can be shared. \cite{googleamzon}. To provide more control on data, data should be encrypted data should be stored. Another use case is:   
       
  %    \item \textbf{All or None Encryption} can we give example of aadhar
 %     \item \textbf{Disk Encryption} Protection of Disk using disk encryption.
      
 \textbf{De-identification and anonymization:} Access control brings restriction on resources of the system. However, personal data is still identifiable to those who have access. To provide further level of protection,  data fiduciary can implement    \emph{anonymization and de-identification} techniques. In this method, when data is collected and processed,  personal identifiable information is removed and replaced with some perturbed data so that data cannot be identified. Translation of identifiable data into de-identifiable form prevents user's personal data from being disclosed.  As per the framework,
 \emph{data fiduciary and data processor shall implement necessary methods such as anonymization and de-identification  during data processing in order to implement appropriate security and safeguard in the system} (as per Article 24.1(a)).  Generally, data fiduciary should anonymize the data before processing, transferring or sharing of personal data to third party. Depending on the anonymity requirement,  different methods can be used  such as \emph{k-anonymity \cite{k-anonymity}, l-diversity \cite{L_diversity}, t-closeness \cite{T_closeness}, differential privacy \cite{differential_privacy}}. Similarly, other methods also exist like, \emph{tokenization, federation, hashing, binning, format preserving encryption       \cite{FPE}, and format preserving hashing\cite{FPH}}. There is always a trade-off between amount of data shared and the method of anonymity that has been applied. 
 \par 
 De-identification process is purely subjective. It completely depends on nature and context of data to decide which method should be used by data fiduciary  in order to implement anonymization \cite{BigDataPbD}. Another assumption is, it is believed that   anonymized data are not really personal data and it can be sold, shared with third parties or freely used without any restriction. However, researchers has shown that in many cases anonymized data can be re-identified \cite{reidentification}. So, protection for both anonymous and unanonymized data are needed. 
 
 %CHECK ref [11] to update journal or conference publication. 

 \subsubsection{Existing limitations of data processing}\label{subsection:limitationDP}
 
 The framework gives excessive power to the central government. For instance, any personal or sensitive personal data can be processed for the function of state or to compliance with law (as per Article 12-14). Similarly, to promote any policy of digital economy central government can direct data fiduciary to provide  any personal data in anonymised form or any other non personal data to create policies or to promote better delivery of services (as per Article 91). Other limitation is, effectiveness of data processing highly depends on transparency in data sharing and disclosure agreement with third parties. 
 
 %For instance, some data fiduciary allows other companies    
 %to puts advertisement on their website. Advertisement company  may have their own data processing policy,  consent and data collection mechanism. If proper agreement of data sharing and disclosure is not settled  between both parties then advertisement company can collect data as per its own preference irrespective of data fiduciary data collection policy. This advertisement company could  collect  personal data even without user's consent. So  necessary requirement  is, both party should have a well define, transparent data collection and data sharing contract.   Advertiser could be  allowed to collect only those data which is as per the contract established with data fiduciary. 
 
 %(give example how this breaching happening, example still it is happening, abd references how it can be removed). 
 \par

 \subsection{Privacy by Design }\label{subsection:pivacybydesign}
  Earlier,  maintaining privacy  was the responsibility of data principal. Software development was focused on technologies  behind which data principal could hide and protect their  data.     Applications  had the aim to make data principal anonymous thus enforcing them to share limited data. TOR, Incognito web browsing, SafeWeb are some  examples to achieve privacy and anonymity from data principal's side. The primary assumption was that data fiduciary  is malicious and  can collect, share and misuse personal information without data owner's  knowledge. The Data protection framework scenario is completely different. Now the responsibility  to protect the privacy of personal data  is shifted to data fiduciary. According to the framework, each data fiduciary should design his system which provides strong privacy of user's personal data and preserve rights of the users. 
   
  %, seven goals of privacy by design has defined, which are: \emph{``Proactive not reactive; preventative not remedial, privacy as the default, privacy embedded into design, full functionality—postive-sum, not zero-sum, end-to-end security—lifecycle protection, visibility and transparency, respect for user privacy''}.
\par 
  Privacy by design (PbD) means that the design of technologies and systems  must protect the privacy of users inherently \cite{PrivacyByDesign}. 
  According to the data protection framework, each data fiduciary has to  implement necessary policies and other activities to accomplish privacy by design (as per Article 22). As stated : \emph{``systems are designed in a manner that avoid harm to data principal, technology used in the processing $\ldots$ in accordance with certified standard, legitimate business interest, and innovation $\ldots$ achieved without compromising privacy, privacy protection throughout life cycle of personal data, transparent processing and interest of personal data shall be accounted''}.  From implementation perspective, the following  will provide a direction to achieve this privacy goal. 
   
  \renewcommand{\labelitemi}{\textendash}
 \begin{enumerate}
      \item Data fiduciary should set a privacy by design objective. He should prepare a \emph{privacy by design architecture} of the system,  so that it describes how  privacy will be  achieved throughout  all the processes in the system. For this, a  design documentation can be prepared and made available to the public. It should be fully  documented. Necessary and major points should be covered about how privacy of users is being preserved. Hiding the design always creates disbelief on data fiduciary \cite{EVTampering}. Therefore, disclosing the privacy design document, so that the data principal, auditor and  third party  can  verify privacy  preservation and data protection techniques in the system.
      
      \item Data fiduciary should implement data consent carefully.  
      (Informed) Consent should include the purpose for collection of data, the entities with whom the data will be shared and how long the data will be stored. If the data is to be shared for other purposes, this should be clearly mentioned. To do this he  can ask more choices, preferences and consent from users. Data fiduciary should implement opt-in methods and check boxes so that user can select and opt for more choices \cite{UninformedConsent}.
      \item 
      Data fiduciary should provide various technical proofs that data is well protected and privacy is being maintained over small to large scale data. Data principles and third party auditors can query the data fiduciary and verify that the privacy features are properly implemented as stated by the data fiduciary. 
      
      %\item The Data fiduciary should disclose a justification why such collection is required. (included in consent)
      \item 
      Data fiduciary should not collect, store, use, share,  and process the data which are not required. 
      
      \item  Data fiduciary should share and  disclose the right data in the right form to the right entity.   If data fiduciary shares data with third parties, it should use privacy enhancing technologies like anonymization or encryption. 
      \item 
      Data fiduciary should maximize the use of encryption methods thoroughly for all processing of data. If processing can be done on encrypted data data  fiduciary must do it using techniques like Searchable encryption, homomorphic encryption and secure multi party computation \cite{Searchable_Encryption, hmomorphic, SMPC}. The data fiduciary should maximize collection, storage and transfer of data in encrypted form. 
      %Apply proper data filtering, anonymization, encryption and other useful safeguard.   

      \item Data fiduciary should provide users more control on their data and  allow data principals to query the present status of data. The data principals should be able to stop  further processing of his data or request to delete his data. 
      
      \item Data fiduciary should provide an option of right to withdraw and tell users why they need it.
      
      \item  As per PDPB,  each data fiduciary should assign a data protection officer(DPO) who will monitor over all privacy and data protection activity of the system and will work as point of contact (PoC) for other parties. 
      
      %They behave a third party auditors. WHAT ARE THE TRUST ASSUMPTIONS? 
      
      \item Data fiduciary should conduct \emph{privacy impact assessment (PIA)} before design of the system is released. He should also  conduct it on regular intervals. Such assessment can be done either by data fiduciary alone or together with data processor. Depending on the nature, scope, size, and kind of data processing, an independent third party may also assigned to perform privacy impact assessment before or during commencement of data processing. Data fiduciary should publish an impact assessment report. 
      
      \item Data fiduciary should inform user about process of data retention, data processing activity, data sharing, data transfer, data collection strategies and data deletion methods uses in the organization. He can disclose and prove  that how he has implemented  privacy of user's data  in significant steps. It will help to build the trust. 
      \item 
      Data fiduciary can also obtain a certificate of the  assessment of  privacy by design activity from any third party audit team approved by DPAI. Disclose this certificate on the website for verification by the users.
      %IS DPAI THE PIA EVALUATOR? PLS CHECK
  \end{enumerate}

 \subsubsection{Cryptographic and other methods to achieve the goals of privacy by design}
  
  %A common trade off is the more information a data fiduciary collect and process more chance will be of privacy disclosure. 
  Use of cryptographic methods can help to achieve a better level of privacy. We discuss below a set of  cryptographic based approach  that can be used to enforce privacy by design. Note that no method is complete. Each method can assure privacy up to a certain degree and have inherent limitations.  Therefore, we have discussed  scenario and  use cases where such methods would be suitable. Subsequently, the benefit and limitation of each method is also provided. Efficient implementation of all these will not only minimize data breach but also provide effective privacy by design based system.
  Though there are other methods to achieve privacy by design like anonymization using k-anonymity, l-diversity, differential privacy, we have not considered these in this paper. For a detailed discussion, one can refer to \cite{WE18}. 
  
  \par
 \textbf{Encryption:} is example of cryptographic techniques which provide both confidentiality and privacy to the data. Using encryption privacy of data can be preserved in both cases when  data is at rest or in motion. Data fiduciary should encourage multiple kinds of encryption to gain higher privacy. However, the limitation of encryption is that the  privacy of personal data can not be guaranteed after decryption of the data or data is available in plain text form. Computation on encrypted data impose communication and computation costs.  
 
 \par
  
  \textbf{SMPC and Homomorphic Encryption:} To achieve privacy further, advanced cryptographic methods could be implemented such as  \emph{secure multiparty computation(SMPC)} \cite{SMPC} and \emph{homomorphic encryption(HE)} \cite{hmomorphic}. Both techniques provide computation over encrypted data. Consider the case of a hospital which regularly share their patient's information with a research institute. Direct sharing of sensitive health data can be risky. Other parties may be able to find out various sensitive information which may result  privacy breach. Privacy preserving  aggregation and computation (of summary statistics) of patient's data could be implemented using \emph{homomorphic encryption} \cite{Gentry_FHE}. To enable data sharing without affecting the privacy, both hospital and research institute can implement \emph{secure multiparty computation}. In SMPC, a set of parties can compute the result function without revealing their data.  Therefore, both hospital and research institute can compute a joint function  through SMPC service upon the health data rather than  sharing data directly.  Using such  methods, data fiduciary/data users can perform computation on the encrypted data  without access of individual's data or secret keys. The resulting computation is also encrypted. The benefit of this can be applied in the area of  sensitive personal data, cloud environment, health data, biometric data or remote computation.   
 
 %Both parties data principal and data fiduciary jointly can do many such computation without disclosing their other sensitive personal like banking or bio-metric data too.
 
 %extract personal inform from the data in may may sharing of health data  Using \emph{secure multiparty computation}  Such case are applicable in  data protection framework where   Since privacy of data is utmost important data fiduciary can use \emph{secure multi party computation(SMPC)} for privacy preserving sharing of data.    

 %Using homomorphic encryption, data principal can store encrypted data remotely and can perform privacy preserving search and manipulation.
 %
 %Google Drive and Dropbox are the famous example regarding this which uses cloud side of key generation and storage of data that  may breach the privacy  by accessing all the data.
  \par
 \textbf{Privacy of Cloud data:} Client-side encryption provides stronger privacy to the users data because key is managed by the user itself. Sending encrypted  data allow users to have ``full control on their data''. However, client-side encryption has limitation that it is useful exclusively when the data is stored on the cloud  and synchronized with the local storage.

  %Privacy risk raises since key is is managed by data fiduciary.  in order to enforce strong privacy. Data fiduciary let allow data principal to encrypt the data before sending to the cloud. It is known as client side encryption.  
  % such storage since key used to encrypt and store the data is managed by data fiduciary. If data fiduciary is malicious all the data can be accessed without prior permission \cite{breach_cloud}. 
  %In this model key is managed by data principal. Storage of cipher text not only  provide security but also strong privacy to the personal data. Several method exists as \emph{``Boxcrypter, Cryptomator''} \cite{Boxcrypter, Cryptomator} which work as man in the middle. In this data is encrypted by the application before sending to the cloud using user's key.    
  
  %the cloud will  achieve security, confidentiality and privacy of user's data as well as it will also achieve the purpose of storing data at cloud. Which is: ``data can be accessed from  anywhere and data principal should
  % Consider the situation where user wants to store the data on the cloud without having any local storage.In future, with the development of advance cloud computation services
  %en encrypted as the above model and stored at remote server.  Since encrypted data is leaving from the client it
   
 \par
  \textbf{Privacy preserving search and query processing:} Suppose, if a data owner wants to reduce the burden of local storage and keep all data to the cloud. Privacy achieved through client side encryption techniques is not sufficient here  because such storage does not allow any meaningful operation like search and query on the data until and unless the completed data is downloaded and decrypted.  This operation is computationally expensive and highly inefficient. Cryptographic techniques such as \emph{searchable encryption} \cite{Searchable_Encryption} and private information retrieval \cite{PIR} can be implemented which enable users to perform query on the stored encrypted data.  
  %In this, a tag value is stored along with encrypted data. Tag value allows users to perform the search query on cipher text.
  %FOR PIR ADD THE REF.  Chor, Benny; Kushilevitz, Eyal; Goldreich, Oded; Sudan, Madhu (1 November 1998). "Private information retrieval" (PDF). Journal of the ACM. 45 (6): 965–981. 
 
  %does not have local storage, wants to store data remotely and data pnot trusted and data principal want to store data remotely.  

 %Disadvantages of this model is, remote server can not fully trusted. In this model, sensitive personal data stored in plain text form  can be  misused. A simple way to achieve privacy is to encrypt the data behttps://www.overleaf.com/project/5c20c35175031d099f5243e5fore sending to the cloud. 

 %and want to  achieve data confidentiality, privacy and remote search. To achieve above data principal stores tag value along with encrypted data. Tag value  help users to retrieve  relevant query. 

\par
\textbf{Provide both confidentiality and anonymity:} There are cryptographic methods  which can achieve further level of privacy such as assurance of both the anonymity of the user as well as the confidentiality of stored data at cloud. It is useful in the scenario where data principal  wants to hide his identity too. For instance, in medical data exchange  program detail of patients(for e. g. HIV patients) needs to preserve the privacy as well as patients would not like to get inferred anything from their identity. Therefore  anonymity is required  along with privacy of user's data.  One such method is \emph{anonymous and confidential encrypted cloud storage} which has proposed  in  \cite{ACECS}. This method protects the identity of user by making data principal anonymous while maintaining the confidentiality of the data.

%For instance, data principal  wants to authenticate himself to the data fiduciary  through some secret information (such as password) but does not want to reveal the secret to the data fiduciary.

\par         
\textbf{Zero knowledge proofs :} More advanced cryptographic method such as \emph{zero knowledge proof (ZKP)} can be used to achieve further level of privacy. Consider the scenario where user wants to get a service without revealing its secret and proving that he posses the secret information. Zero knowledge proofs are useful in such cases. Using this, any user can prove their \emph{honest and ethical behavior} while maintaining the privacy of his secret information. In data protection framework one example is \emph{children age verification} (as per Article 16) where it can be used. Any child can prove his right age in order to get a service without revealing the exact age or date of birth. On the other side, data fiduciary can use it to verify the fact that child has valid age the knowledge real data of birth is not necessary.  

\par
\textbf{Blockchain technology:} Another way to achieve privacy of personal data is through integration of  \emph{blockchain technologies}. It is a good alternative  for privacy preserving data sharing, data exchange  and data market place. Since the inherent property of blockchain is decentralization, privacy and transparency   both data principal and data fiduciary can implement it to achieve privacy by design. 
For instance, consider a user who wants to find a match on an online dating service without revealing the
criteria of finding partner to anyone (not even to third party who stores the data set). The third
party will return the output to the user after obliviously searching in its data set. This can be
achieved using the combination of  searchable encryption technique and blockchain technology that allows a party to perform oblivious match on the data without revealing the input \cite{bc578_matrimonial}. Another example is \emph{zero knowledge contingent payment(ZKCP)} protocol that allow fair exchange of goods and payment without any third party \cite{ZKPcontingentPayment}. 

\par 
Privacy of data is completely subjective. It depends on nature and context of the user's at which extent he wants to achieve the privacy. Data fiduciary can analyze  the nature and the scope of processing, severity of harm and can implement different technologies  to facilitate  suitable privacy to the  user's personal data. Such implementation also  depends upon where and how much data has been disclosed.  

 \subsection{Transparency} \label{subsection:transparency}
 
  Transparency can be achieved through openness and accountability. Article 23 of data protection framework  states: \emph{``the data fiduciary shall take reasonable step to maintain transparency...''}. To achieve it, data fiduciary may avail various kind of information easily accessible like \emph{``category of personal data collected, purpose of collection, the existence of procedure.. for data principal rights.., platform to lodge complaints, information regarding cross border transfer.. , or any other information'' }. 
 
 \par
 A transparent system builds more trust among users. To gain more confidence of users, data fiduciary should keep more transparency in their process. To achieve this, he can disclose  multiple
  number of verification, validation and check points to data principal. The check marks could be their processing activity, design of system, the way security of data is being implemented, system security architecture, or the way privacy is being preserved within or outside the system. A trade off always exists between privacy and transparency. It is a fact that more number of disclosed parameter leads to the  reduction of  security and privacy  but on the other side  openness is also needed to achieve the transparency at a greater extent. Therefore, it is purely subjective to data fiduciary how he can establish coordination between both and builds trust among users. 
  
  \par 
  We discuss set of properties that  could be made available to the users to make processing more transparent. These points provides a direction how data fiduciary can achieve transparency as per the framework.
  
  \begin{itemize}
      
     \item  Disclose high level security design of system.  Almost every organization claims that it uses best practices to secure the system still breaches are happening. To gain trust of people it is required to disclose security design of system so that the users can ensure and verify that the personal data is being secured.  For instance consider the usages of \emph{electronic voting machine(EVM) \cite{EVM}}. Govt of India  demonstrate its security feature periodically and confirm that it is fully secured. Still, people  suspect the security of  EVM and always challenge that it  can be tampered \cite{EVTampering}. Disclosure of  security design by data fiduciary can provide more transparency. Making design public will build more trust on the system as user can validate its security parameters against various attacks.
     
     \item Disclose list of parties with whom data fiduciary is sharing personal data. This list will help users to track their data. It is required because list of third parties are completely black box till now. Currently, policy of most of the server does not  disclose the list of parties with whom data is being shared. 
     
     \item Disclose whether data fiduciary implements any encryption methods. Also, disclose the kind of encryption and different level of encryption methods have used.
     
     \item Disclose whether data fiduciary implements any access control techniques. Who has access of what? For instance data fiduciary can announce that system provides geo-location based access control. 
     \item Data fiduciary should make advertisement process more transparent with respect to user.
      
      \item Disclose to data principal why data fiduciary is collecting data. List of category of personal data and the procedure using which data is being collected.
      
      \item Inform user's what are their rights  and how they can achieve it. Implement web pages easily accessible to lodge complaints for right to access, right to delete etc.

      \item  Data fiduciary can disclose the contract using which data is shared with third parties. Security and safeguard of such data should be bound by such contract. 
      \item Disclose if there is any cross-border transfer of data. If yes, disclose whether explicit consent is taken or not. 
      \item Disclose whether system implements anonymization and de-identification technique. 
      \item Disclose the security standard data fiduciary follows. Data fiduciary may disclose the architecture of their security within an organization. It may help data personal to understand security while sharing his data. For example, Dropbox provides high level security description \cite{DropBox_Security} of  storage of data on the cloud.    
      \item   Disclose trust level of your organization. It can be done be getting certificate from protection authority.  
      \item Disclose rating of the organization publicly obtained from the authority.
      \item  Data fiduciary shall notify the data principal of important operation in the processing of personal data. Data fiduciary can announce how frequently audit/checks is performed in order to ensure privacy and security safeguard.
      \item Disclose how code of conduct is followed with in organisation. 
      \item Disclose the audit procedure. Nature and scope of audit can me made available. Disclose how many kind of audit the system implements. 
      \item  Disclose data collection method used by data fiduciary.
      \item Disclose data retention and deletion period. Disclose data deletion method used and proofs of honest in the deletion.
      \item Disclose Privacy be design method. What are the operations used by data fiduciary to achieve the privacy. 
      
      \item  If data fiduciary is providing storage services then  data fiduciary can disclose about \emph{proof of storage, proof of space, proof of location, proof of possession of data, proof of retrievability} methods.
      %\item Disclose if there are any cross border transfer of data.
      \item Data fiduciary can disclose Data Breach Mitigation Plan (DBMP) . It can also disclose how much transparently breach notification system is established with the authority.
      \item Disclose whether system has CISO, security handling team, data protection officer, and impact assessment team. It will help to report data breach, to resolve query and other incidents of data protection.
      \item Disclose whether system implements more advanced method of security and privacy such as data loss prevention(DLP) \cite{DLP} methods, insider threat detection and prevention(ITDP) \cite{InsiderThreat} or security operation center(SOC) \cite{SOC}. 
  \end{itemize}
  
 %can we use blockchain to achieve transparency??

 \subsection{Data protection audit, rating and data protection impact assessment } \label{subsection: audit}
 
 Data processing  will be scrutinize by an auditor. The purpose of audit is to verify that system achieves the intended goal of the data protection. Auditor is an independent third party registered with authority who may conduct the \emph{audit}. As per Article 29, \emph{``data fiduciary shall perform, data audit annually..., the audit will evaluate, effectiveness of consent, privacy, transparency, safeguard and any other kind of audit which may be specified"}. 

 \subsubsection{Challenges in data audit and rating}
  Defining audit procedure in the context of data protection framework is highly important  because \emph{``it will help  to verify the proof and correctness of many features such as  data collection, data consents, data storage and data deletion.''}.  
To keep higher transparency, the audit should  minimize manual auditing method and should encourage automation. As per the framework, it is the responsibility of data protection authority to specify the parameter and procedure for data protection audit. However, currently no such specification exists. Therefore, we first point out the list of challenges that needs to be considered for the specification and implementation of data audit procedure.
  \begin{itemize}
      \item How audit should be conducted and what will be the procedure, parameter and set of criteria to design audit and empanelment process.  
        \item How the audit conducting  authority  will implement the procedure of rating?
      \item How to verify that the rating disclosed by an organization is correct and generated from verified auditor?
      \item What are the period when audit can be conducted such as six month, yearly etc. 
      \item What are the techniques using that audit can be automated with minimum manual interference?
      \item In some cases auditor may be malicious and he can submit false report. How transparency in the audit can enforced when the auditor is malicious?
      \item  Should data fiduciary also  provide any procedure of audit to data principal?
      
\end{itemize}
 Currently standards for security audit already exists \cite{NIST_audit}. These standards primarily assess information security risks \cite{certin_audit} and, the auditor generally conducts  vulnerability assessment, penetration testing, review of used security standards and  verification of access control policy. However, there is no standards for data protection audit. One way to design audit standards is by modifying the standards of security audit. The resultant broader framework would contain all the steps of security audit as well as many additional steps of data protection audit. It can be called as {\emph{data protection audit framework}(DPAF)}. It should contain parameters, standard operating procedure and techniques for audit. It will be a procedural  guideline by which  auditor can perform verification and validation. We now describe what kind of proofs authority can includes in DPAF. 
 %We have  discussed a set of  that can used to build the data protection audit framework(DPAF).

\subsubsection{Design of Data protection audit framework (DPAF)}
DPAF should define the technical specification, form, manner, and standard operating procedure clearly through which the audit will conducted. Many of the  audit procedure will require advanced technical implementation . We suggest the following proofs and validation as an example which can be included in the data protection audit framework (DPAF). These are not final, many other properties can  be explored, formalized and included in the DPAF.

  \par 
  \textbf{The auditor can ask consent management proofs:} Auditor should check how data fiduciary establishes consent with data principal. How consent are updated and managed? He can also review the list of information data fiduciary disclose for the consent establishment. Next, he can also investigate what are the  additional information data fiduciary is adding along with consent. This could be a list of third parties or data processors with whom personal data being shared, disclosure of data retention period, cookies policy or information about any cross border transfer. Further, it could be  also checked how the \emph{consent manager} is being implemented.
  \par 
  \textbf{The auditor can verify compliance of PDPB law:} The Auditor should  verify whether the design of the system is compliance with PDPB law?. Means whether data fiduciary follow purpose limitation, collection limitation, quality control, accountability or other conditions of processing. The auditor can verify  whether they are doing any cross transfer of data or not. For instance, the auditor can ask geo-locations proof of the data.
  \par 
  An assessment can be done to examine the data retention policy. The auditor can verify  how long data is  being stored. Does data principal get any notification about the retention period of data? Whether data fiduciary does any periodical review of retention or not? Data auditor can ask whether any cryptographic proof for retention of data such as  proof of storage or proof of retrievability is used? Data fiduciary can also give the proof of concept of how right of erasure request is handled when it is raised by data principal. On the other side, data auditor can check whether data fiduciary implements any proof of erasure technique to prove the deletion.  Subsequently, it may also verified, what are the data storage location, storage back up plan of personal data. Further level of check can be done to confirm that how data is being erased from multiple geo-locations.   
  
  \par
  \textbf{The auditor can ask \emph{proof of X}:} It can be the list of proofs of various techniques that data fiduciary has implemented to protect data processing. It may include proof of concept such as  proof of storage \cite{ProofOfStorage} whether the remote server has indeed storage as claimed by data fiduciary, \emph{proof of space} \cite{ProofOfSpace} in order to check that stored data has not been modified and changed. Similarly, auditor can check whether data fiduciary provides \emph{proof of location} to ensure what is the list of all locations where personal data reside, \emph{proof of possession} to verify data fiduciary indeed holds user's personal data. Further, the auditor can also verify  \emph{proof of retrievability} to know whether stored data can be retrieved or not or  \emph{proof of geo-location} to check is there any cross border transfer of data.  It may be possible that all proof may or may not be implemented. But the goal of data fiduciary should be to maximize the number of proofs in order to review of transparency of  the system.
  
\par
Data fiduciary who is processing children data, the auditor can verify what are the age verification mechanism being used, how parental consent is established. If the data fiduciary is guardian data fiduciary, auditor can confirm whether data fiduciary does any  profiling,  behavior analysis, target based adverting or other acts which can harm to the children.  
\par
\textbf{The auditor can verify security safeguard methods:} An evaluation can be done to verify that what are the  security and safeguards implemented within the system. What are the encryption techniques data fiduciary is using, how many layers of encryption has done, list of  authentication, set of authorization techniques data fiduciary uses to protect the data? Whether the system implements any de-identification and anonymization method. What are the access control policy of the  data fiduciary? Similarly, auditor can verify methods used to implement privacy by design and transparency. 
\par
\textbf{The auditor can verify data distribution procedure:} Further level of verification can be done regarding the mechanism used for  distribution and selling of data. To verify this \emph{sell and purchase agreement} can be demanded. Auditing team can also request information regarding how data fiduciary handle \emph{data breach}. What are the mitigation plan  available in the system to prevent and minimize data breach and the data loss. Additionally, it can be verified that how data fiduciary communicates data breach and other things to the authority. Subsequently, a list can be obtained regarding last few breaches happened in the system (if any)   and the steps that  have been  followed to remove them.
\par 
 Data fiduciary should  also conduct audit periodically by itself to verify the correctness of processing internally. This kind of audit is called as \emph{data protection impact assessment}. In such assessment he can follow the \emph{the role reversing} method. For this, data fiduciary can consider himself as data principal and audit the system from his perspective. All data protection requirement from the view of   data principal can be listed and examined by applying  all the procedure of data audit. Using this, data fiduciary can understand the complete security and privacy  design of the system. 

\par
All the above use cases will help to  build the data protection audit framework. A proper design can be done for each points and a strategy can be developed how it will be implemented technically at the ground level. Against each checkpoints a weighted score can be assigned. Based on overall score a \emph{rating} can be generated and a \emph{certificate} can be provided to the data fiduciary. Data fiduciary can publish this certificate on his website for user awareness.
    
\subsubsection{Cryptographic techniques for data protection audit framework (DPAF)}

The audit should minimize human intervention. More manual intervention may lead to prejudice in the audit procedure. For instance, the goal of audit can not accomplished if few parties are  malicious. We discuss here how cryptography techniques can provide consensus in data audit even if few parties are malicious. We discuss here two such methods.

%and  blockchain based data audit. First one ensure consensus in the  audit when data fiduciary can be malicious and auditor is honest. In the second case where party one can be malicious, blockchain based auditing method can be used. A brief discussion ob both is given below.

\par
  
  Audit can reveal sensitive information to the auditor. If the auditor is not honest then such data can be used maliciously. Therefore, audit should performed on personal data without influencing the privacy.  \emph{Privacy preserving data audit} \cite{TPA1,TPA2} is the example of such techniques which allow  an independent third party auditor(TPA) to perform the audit on the  stored data at remote server without affecting privacy.   
     %are the examples of such model that provide privacy preserving data audit and proves the correctness of data without affecting privacy of user's . 
    
    Further, the above model has limitation that it assumes data owner and the third party are honest and data fiduciary could be malicious \cite{TPA1}.  There are
     \emph{blockchain based privacy preserving data audit \cite{BDA1, bc10_1_DataStorageIoT}} \label{subsection:blockchaindataaudit} models present in the literature which can implement data  auditing even if all the parties (data owner, data fiduciary and auditor) are malicious. In such settings, blockchain helps to provide consensus in the audit.

    %Privacy preserving data audit using     has proposed blockchain based privacy preserving data audit. In this model, it is assumed that any party can be malicious.  

  \par
  The ultimate goal of the audit is to evaluate overall system activities and make an assurance  that data processing is up to the standard of data protection framework. The auditor can provide a \emph{rating or trust score} to data fiduciary. This certificate  will show how much processing is compliance with data protection framework. 
  
  %To  bring more transparency, trust and to minimize prejudice in the audit data principal can be present at the time of the audit. Alternatively, data fiduciary can handover some part of verification at the end of data principal too.  Such as disclosure of few designs (such as security design, architecture of privacy by design) publicly so that data principal  can also validate it.

  %Generally auditor will  be a third party registered at DPAI. Auditing may reveal several sensitive information about the  system. To prevent this, data fiduciary should promote privacy \emph{preserving data audit technique} which shall not leak any information. 
  
  %\par 
  %that can be used by auditor to conduct audit, to generate trust score and to provide rating. we also discuss what check points that need to be consider in order to make audit more transparent and trustful.

 %\subsubsection{Rating and trust score}
 %As per Sub Section 35(5), \emph{"Data fiduciary may assign  a rating in the form of trust score"}. A score or certificate can be provided by data auditor which will display trust and confidence level of organization.
 
 %\subsubsection{data protection impact assessment}
 %While designing 

 %\subsection{Processing with fair and %reasonable with lawful manner}
 %data collection data processing and data %deletion should be achieved without affecting %privacy . can be achieved through crypto %techniques.
 
 \subsection{Assurance of users rights}\label{subsection:userright}
 Data principal has \textit{right to access}, \textit{right to correction, right to erasure} and \textit{right to forgotten} of their personal data as per Article 17, 18 and 20 respectively. Data principal can ask in a plain text form the set of collected data, the list of data generated from processing and  methodology used for processing of personal data. Data fiduciary has to provide  a method using which correction and update  in personal data can be recorded as per the request of data principal. User's have right to  prevent data fiduciary from further processing of personal data as well as disclosure of data  to third parties. 
 \par
  In the real world, rights of access as well right of correction can be provided to the users easily. For this data fiduciary can display a web portal where data principal can  request necessary update and correction. The bigger challenge may arise at the time of dispute between both parties. For instance, data principal finds that  correction has not recorded, or  processing is being done as per the old data even after  correction has been updated. Data fiduciary has to provide a proof that  data  has accessed, or a proof that  correction has recorded and processing is being done on updated personal data. To prove all these, a \emph{record manager such as Blockchain} \cite{blockchain} can be used that can record and prove what data has accessed, corrected and updated. The implemented method should also able resolve the dispute. 
 \par 
 \textbf{Proof of deletion and immutability of data:} Data principal has right to erasure and right to forgotten. In the first case user can do the request to delete their personal data while in second case user can prevent  data fiduciary from further processing and disclosure of data to the third parties.  In the real world it is  difficult to prove honesty of data fiduciary in both right to erasure and right to forgotten. It is hard to prevent data fiduciary from further processing and  disclosure as well as it is also challenging to verify that data is really erased  due to various technical limitations \cite{7sin}. Further data processing community has no consensus on deletion of data. As in \cite{EfficacyOfGDPR}, author has discussed about the efficacy of right to forgotten tenet of GDPR. They mentioned that right to forgotten is effective thing, but business model of many organization like Facebook rely on  user's data only. Request of deletion of personal data may affect the existence of their business model. Similarly, personal data is also processed by  blockchain \cite{bc10_1_DataStorageIoT, BuisnessOfPersonalData} . Due to immutable property of data in blockchain it would not be possible to delete the data stored in the chain. By considering the efficacy of deletion of data in GDPR, \cite{BlockchainMutability} has discussed about this immutability problem and  mentioned challenges and possible solution to delete the data from blockchain. Overall deletion  of data in the real world is still challenging task.
 
 \subsection{Data breaches}\label{subsection:databreach}
  Data breaches are the incident due to disclosure of personal data intentionally/unintentionally.  In the framework  it is stated as"\emph{``any unauthorised, accidental disclosure, acquisition, sharing, use, alteration, destruction,...that compromise confidentiality, integrity or availability of personal data to data principal"} is known as data breach. \emph{``Any data breach shall be notify to the data protection authority"} (as per Article 25 ).  It is essential to report the data breach to the authority.  Depends on the sensitivity of the data, authority may decide whether  breach should be reported to the data principal or not. 
 \par
 \textbf{Necessity of transparency: } Data breach of an organization can be reported either internally or externally. Therefore, the essential technical problem is, how transparency can maintain during the reporting of the breach?  If external entity reports the breach it would come under the observation of data protection authority immediately. But if the data breach is identified internally then transparency in the reporting would entirely depend on the honesty of data fiduciary. It may possible that data fiduciary can hide it and does not notify to the authority. In such cases, the consequences can not be determined. This can be seen in the recent data breaches also where organization has not accepted any loss of personal data even after reporting of the breach \cite{cambridge}. The other reason why transparency is necessary because sometimes organization accept about the breach but they strongly deny the  loss of any personal data \cite{Microsoft_million}.  However, no technical solution exist that can ensure transparency in the reporting of the data breach. From legal perspective, a penalty is specified in the framework which can be imposed if the authority finds a data breach incident concealed by data fiduciary. 
 \par
 \textbf{Unfairness in the reporting:} Reporting of the data breach is not fair. PDPB makes  reporting  very restrictive by isolating data principal from breach notification. Personal data belongs to data principal, therefore, data breach should be also reported to the data principal along with DPAI. Further, if data breach has the impact of \emph{high risk} authority may direct  data fiduciary to inform about breaches to the data principal. During such reporting, he  has to  provide guidelines about what remedial actions have been performed to mitigate the risk. Presently, the definition and the criteria of high risk is not specified in the bill. DPAI should publish a definition of \emph{ the type and nature of the risk} which can be considered as high risk.
 
 %in some cases based on severity of harm and th to the personal  breaches will be reported in the case of ''high risk" only. 
 \par
 \textbf{Implementation:} From implementation perspective detection and validation of data breach can be done through   \emph{data protection audit} and \emph{data  protection impact assessment} of the system.  A third party independent audit team can also be assigned for the quest of data breach. The team can proactively observe the presence of breach (if any). To do this, the audit team can examine and verify the sensitive operations such as  modification, disclosure, manipulation or deletion of the personal data. Other verification such as access control or authentication can be also checked. Data fiduciary can also implement more advanced incident handling procedure such as \emph{threat hunting} \cite{ThreatHunting}  which can look continuously for the presence of threats that could create a data breach. Further,  \emph{data loss prevention} \cite{DLP} can be implemented  to prevent data loss from vulnerabilities.  Many times data breaches happen due to \emph{insider threats} \cite{Amazon_insider_databreach}. To prevent this, data fiduciary can execute \emph{insider threat analysis} program in their system periodically or could implement \emph{insider threat management} plan \cite{InsiderThreat}.

 \subsection{Processing of sensitive personal data and critical personal data} \label{subsection:localisation}
 Data fiduciary has to inform to the data principal if there is any cross border transfer of personal data. He can inform him at the time of consent establishment.  For \emph{sensitive personal data} then restriction becomes more stronger. Sensitive personal data can transferred out side India if authority approve such transfer (as per Article 33). Approval is to be determined on the grounds of type  and nature of sensitive personal data, international relation with other country or any other condition as specified. Additionally, data fiduciary has to maintain a local copy of sensitive personal data  within India. 
 \par
 Further, data protection framework restrict that \emph{critical personal data} can be stored and processed within India only. Any data fiduciary who is providing service to Indian customer requires to install their gateway and application services located within Indian territory. Storage of a local copy  personal data within India will increase hardware and software cost as well as the  man power. Ultimately it will increase overall management cost of the system. 
 
 \par 
 \textbf{Why localization is necessary:} It is still controversial why PDPB has such restriction? As per our point view the main reason might be that India government wants to hold more control on  how data should flow across the border. Here with following reasoning we argue that such restriction have been kept purposefully. 
 \begin{itemize}
     %\item Consider a DNS service provider which host malicious website which is hosted outside the country. This Accessing and blocking malicious website. 
      \item To achieve accessibility of the data \emph{to pursue the law}. For instance, to maintain the law  state may require data from an organization who provide service to the Indian customer. But, if the storage is located out side the border it can be denied as per the privacy policy of foreign country or the data sharing agreement between both country. Localization of data will help to access these data easily in order to pursue the law.
     \item Indian government is more concern about the  geo-location availability of the data in addition with  protection of sensitive personal data and critical personal data.
     \item To conduct data audit easily for protection of the data. Authority can do the  data protection audit freely and to ensure safety of data. 
     \item It will ensure the availability of data to analyse data breach  from recent attacks. Ease of access and the availability would also assure that data can be used for  other purposes such as  statistical analysis, research , innovation and policy making.
     \item Government's data are highly critical. A strong  security protection is required while processing. Availability of this data to other country might  have significant risks. For instance, to implement cloud services  for data processing  it is a debatable point whether storage of cloud should be  with in India and government should  follow the policy of localization of data or not?  These  data are sensitive and cross border transfer could be harmful. \cite{cloud_indiadecision}. 
     
 \end{itemize}
 
 From implementation point of view, if local copy of sensitive personal data is stored within India then a proof  will be required to prove the geo location of data. Also, data fiduciary has to prove that local copy of India is complete, accurate and updated. 
 
 \subsection{Processing of children data}\label{subsection:children}
 It is a common belief that children are easy target over the internet. Therefore, PDPB restricts that children data will be processed with higher protection and  data fiduciary has to preserve the rights of children. To accomplish this, each data fiduciary has to implement an \emph{age verification mechanism} in order to verify the age of the children. Further, data fiduciary should incorporate \emph{parental consent} mechanism to acknowledge for parent consent (as per Article 16).   Additionally, the authority will publish \emph{the list of websites who are engage in the processing of children data} . Data fiduciary has to disclose  the nature (such as providing counselling and child protection services) and the  volume  data being processed. Based on it, DPAI will announce some data fiduciary as \emph{guardian data fiduciary}. They will be \emph{"barred from profiling, tracking, behavior analysis and target based advertisement"}. 
  
  %An \emph{``appropriate mechanism for age verification and parental consent shall be incorporated.."} 
 \par
  \textbf{Age verification:} From technological implementation perspective the data fiduciary has to  implement a system for the verification  of children's age. In literature few work exists regarding it, such as an online anonymous age verification system has been proposed in \cite{Anonymous_age_verification}. Any individual can obtain an universal age identification number(UAID) from a legal authority.  This number can be used to verify their age on the website to get  controlled access. Similarly, other techniques such as  such as \emph{anonymous credential \cite{anonymous_credential}, zero knowledge proof based verification \cite{ZKP}} can be useful to prove the age.
 \par 
 \textbf{Parental consent:} Internet is a power full platform for the children to get education, gaming, sports and the entertainment. While on the opposite side they are big opportunity for online market. Children can be easily targeted and can be the victim of privacy breach, harassment, and disclosure of sensitive personal information. To prevent this, monitoring and notification of children activity to the parent can provide an appropriate safeguard  to the children. Practically, many  \emph{parental control} methods exist which enable guardians to control children online behavior \cite{parental_consent_safeguard}. Use of such regulation enable children to accomplish the effective use of internet and  prevent them being a victim online market \cite{parental_consent_harasment}.  

 \par 
 \textbf{Challenges:} The major challenge  is to categorize the list of service providers who is processing children data. Next, how such list will be periodically updated? The another challenge is, how  will be verified and validate that  profiling, tracking, behavior analysis and target based advertising is not being done by the restricted data fiduciary?  
 
\section{Major Challenges 
and Limitations of PDPB}
The data protection framework is a major step towards the protection of individual's data. However, it is  still not clear  how adequately it will achieve its objective. This uncertainty arises due to both the inherent limitation of data protection bill and the constraints of practical  implementation.   Here we discuss some limitations and challenges present in the framework which needs to be taken care:

%(TO ADD THE RELEVANT SECTIONS/ARTICLE NUMBERS FOR EACH OF THESE)

\par
\textbf{Disclosure to the government:}  This bill provides \emph{excessive power to the government} for processing of personal and sensitive personal data. Boundary of privacy is not defined when state processes the data.
Government can ask any personal data for the functioning of state, during emergency, for the security of the state, for the purpose of prevention, detection, investigation or prosecution of any offence, or for any other contravention of the law (as per Section 12).  Further, to promote any policy of digital economy, the government can direct data fiduciary to provide  any personal data in anonymised form or any other non personal data to create policies or to promote better delivery of services (as per Article 91). Request for such kind of data is controversial  and  gives unlimited power to the government.

\par
\textbf{Cloudiness in definition:}
Data protection framework will create a regulatory body known as data protection authority of India (DPAI). It has an independent regulatory power to monitor the data protection activity. It has control to update and approve various definitions, capability to provide notification and direction to the data fiduciary. Further, it has rights of call to information, rights of search and seize, rights to conduct the enquiry and has power to coordinate with other regulatory authorities. PDPB has many undefined terms that will be defined by DPAI in the future. For instance, definition of critical personal data, list of guardian data fiduciary or social media data fiduciary has to be specified. Similarly, the boundary between  \emph{personal data, critical personal data and sensitive personal data} is not yet defined. Further, a certain category of biometric data will be prohibited from processing (as per Article 92). The name of such specific category is not yet defined but later it will be notified by the DPAI. Subsequently, any breach in the personal data shall be reported only to \emph{DPAI}. In the case of  high risk, it can be reported to the data principal. As of now,  definition of \emph{high risk} is not mentioned. 
\par 
Further, the  exact motive behind the localization of at least one copy of sensitive personal data and complete localization of critical personal data
  is still not clear.  It may be either for maintaining control on the data or to ensure \emph{availability of data}.  If it is regarding availability  then many other technical standards also exist which can provide stronger availability than maintaining a local copy of data. 
\par
\textbf{Requirement of fairness:}
  \emph{Verification of distribution and reselling of data} will be very challenging in the era of data market where every one wants to sell data.  The other challenge is how transparency will ensured in 
  \emph{certificate, rating and trust level generation}? It is not clear how framework will coordinate with data protection regulation and law of other countries? For instance, PDPB asks to  assign a local representative if company does not have an establishment in India, but providing service in India. However,  it is not clear how the incidents will be handled if data policy of one country overrides the others?
  
  \par 
%   \begin{comment}
%   \textbf{Incentives and penalties:}
%   Penalties and compensations are defined under Article 57-66.  Different penalty can be imposed based on the nature of violation under the regulation. One instance is, if data fiduciary  fails to comply with the appropriate security safeguard or violate the condition of children data processing then he/she may have to pay ``up to 15 crore rupees or four percent of its total worldwide turnover of preceding financial year whichever is higher''. 
%   But it is not known, how violation of regulation and non-compliance will be proved?  This is proved in court of law. So commented this. 
%   Another  instance is, there is a penalty if  data fiduciary violates the clause of cross border transfer. But, to verify and prove such misconduct would be a challenging task technically. Similarly, how the reporting, response and verification of such complain would be validated fairly?
%   \end{comment}

%\begin{itemize}
  %\item \Data protection authority of India
 % \item 
 
            %\item The criminal liabilities making all offenses \emph{cognisable and non-bailable} under this bill. 
   
  %  \end{itemize}

  %\item  How DPAI will \textbf{work with other collaborative organizations} like CERT-In,Cyber, NIA, CBI etc.??
    %\item \textbf{Each website is taking user’s data browsing experience, location information etc Just to enhance browsing experience? Is this really required? (what restriction should be on this activity to restrict unnecessary personal data) is this covered in protection law?? }.

\section{Conclusion}
 
After the introduction of PDPB, the major challenge will be to translate legal tenets into technological implementation using existing technologies. In this paper, we have discussed the Indian data protection framework (PDPB) in details from  technical perspective. The high-level idea for  implementation of each major obligation has been presented. The various features like data consent, data collection, data processing, privacy by design and data audit has been explained thoroughly. A set of cryptographic and non-cryptographic solutions along with various examples and use cases have been also been provided. This high-level idea of methods for legal tenets can motivate developers to use it for the development and building the real world system as per the PDPB standard. Along with technical implementation, we have also discussed existing challenges and limitations that will require to solve to make data protection framework more strong. Overall, this work will provide a direction to the software developers, system designers, organisation, and data fiduciary to design their system as per PDPB by  developing and implementing PDPB standard based technology at a different levels. This paper will also work as a guideline for other  regulatory authorities to change and set set their activities as per the new  standards. In future work, more advanced methods can be developed to solve the existing limitations. A new set of techniques  will be  explored for more effective implementation of major tenets like data consent, privacy by design and data collection.       

%\bibliographystyle{plain}
%\bibliographystyle{acm}
% \bibliographystyle{ieeetr}
% \bibliography{link}

\end{document}